\documentclass[final]{cvpr}
\pdfoutput=1
\usepackage{enumitem}
\usepackage{amsmath}
\usepackage{times}
\usepackage{graphicx}
\usepackage{bm}
\usepackage{amsfonts}
\usepackage{multirow}
\usepackage{color}
\usepackage{booktabs}
\usepackage{ifthen}
\usepackage{gensymb}
\usepackage{mathtools}
\usepackage{hyperref}
\usepackage[capitalize]{cleveref}
\usepackage[normalem]{ulem} % for sout
\usepackage{array} 
\usepackage{multirow}
\newcolumntype{C}[1]{>{\centering\arraybackslash}p{#1}}

\definecolor{gray}{rgb}{0.5,0.5,0.5}
\definecolor{green}{rgb}{0, 0.6, 0}
\definecolor{orange}{rgb}{1, 0.5, 0}
\definecolor{mahogany}{rgb}{0.75, 0.25, 0.0}
\definecolor{purple}{rgb}{0.6, 0, 0.6}
\definecolor{darkgreen}{rgb}{0, 0.3, 0}
\definecolor{orange}{rgb}{1, 0.5, 0.}

\def\cvprPaperID{666} % *** Enter the CVPR Paper ID here
\def\confName{CVPR}
\def\confYear{2022}

\usepackage{xparse}
\usepackage{multirow} 
\NewDocumentCommand\pquotes{ m g }{%
    \IfNoValueTF{#2}{% 
         \textit{``#1''}%
    }{%
         \textit{``#1''}{\,}{\small(#2)}%
    }%
}
\begin{document}

\title{PreCare: 
Designing AI Assistants for Advance Care Planning (ACP) to Enhance Personal Value Exploration, Patient Knowledge, and Decisional Confidence 
}
\author{
Yu Lun Hsu\\
National Taiwan University\\
b06401078@ntu.edu.tw\\
\and
Yun-Rung Chou\\
National Taiwan University\\
b11705023@ntu.edu.tw\\
\and
Chiao-Ju Chang\\
National Taiwan University\\
r11922057@csie.ntu.edu.tw\\
\and
Yu-Cheng Chang\\
National Taiwan University\\
r11922021@ntu.edu.tw\\
\and
Zer-Wei Lee\\
National Taiwan University\\
r12922205@ntu.edu.tw\\
\and
Rokas Gipiškis\\
Vilnius University\\
rokas.gipiskis@mif.vu.lt\\
\and
Rachel Li\\
University of California, Berkeley\\
rachel.l@berkeley.edu\\
\and
Chih-Yuan Shih\\
National Taiwan University Hospital\\
oscarpetershih@gmail.com\\
\and
Jen-Kuei Peng\\
National Taiwan University Hospital\\
jkpeng@ntu.edu.tw\\
\and
Hsien-Liang Huang\\
National Taiwan University Hospital\\
tennishuang@gmail.com\\
\and
Jaw-Shiun Tsai\\
National Taiwan University Hospital\\
jawshiun@ntu.edu.tw\\
\and
Mike Y. Chen\\
National Taiwan University\\
mikechen@csie.ntu.edu.tw\\
}

\twocolumn[{%
\renewcommand\twocolumn[1][]{#1}%
\maketitle
\vspace{-1cm}
\begin{abstract}
Advance Care Planning (ACP) enables individuals to document their preferred end-of-life life-sustaining treatments prior to potential incapacitation due to injury or terminal illnesses such as coma, cancer, or dementia. While online ACP platforms offer high accessibility, they often lack essential benefits provided by clinical consultations, including deep introspection of personal values, real-time Q\&A on medical treatments, and personalized reviews of decision consequences.  
To bridge this gap, we conducted two formative studies: 1) shadowing and interviewing 3 ACP teams consisting of physicians, nurses, and social workers (18 patients total), and 2) interviewing 14 users of ACP websites.
Leveraging these insights, we developed PreCare in collaboration with 6 ACP professionals. PreCare is a website featuring 3 AI-driven assistants designed to guide users through exploring personal values, gaining ACP knowledge, and supporting informed decision-making.
A usability study (n=12) showed that PreCare achieved a System Usability Scale (SUS) rating of excellent.
A comparative evaluation (n=12) showed that PreCare's AI assistants significantly improved exploration of personal values, knowledge, and decisional confidence, and was preferred by 92\% of participants. 
\end{abstract}

}]
\twocolumn[{%
\begin{center}
    \centering
    % \captionsetup{type=figure}
    \includegraphics[width=\linewidth]{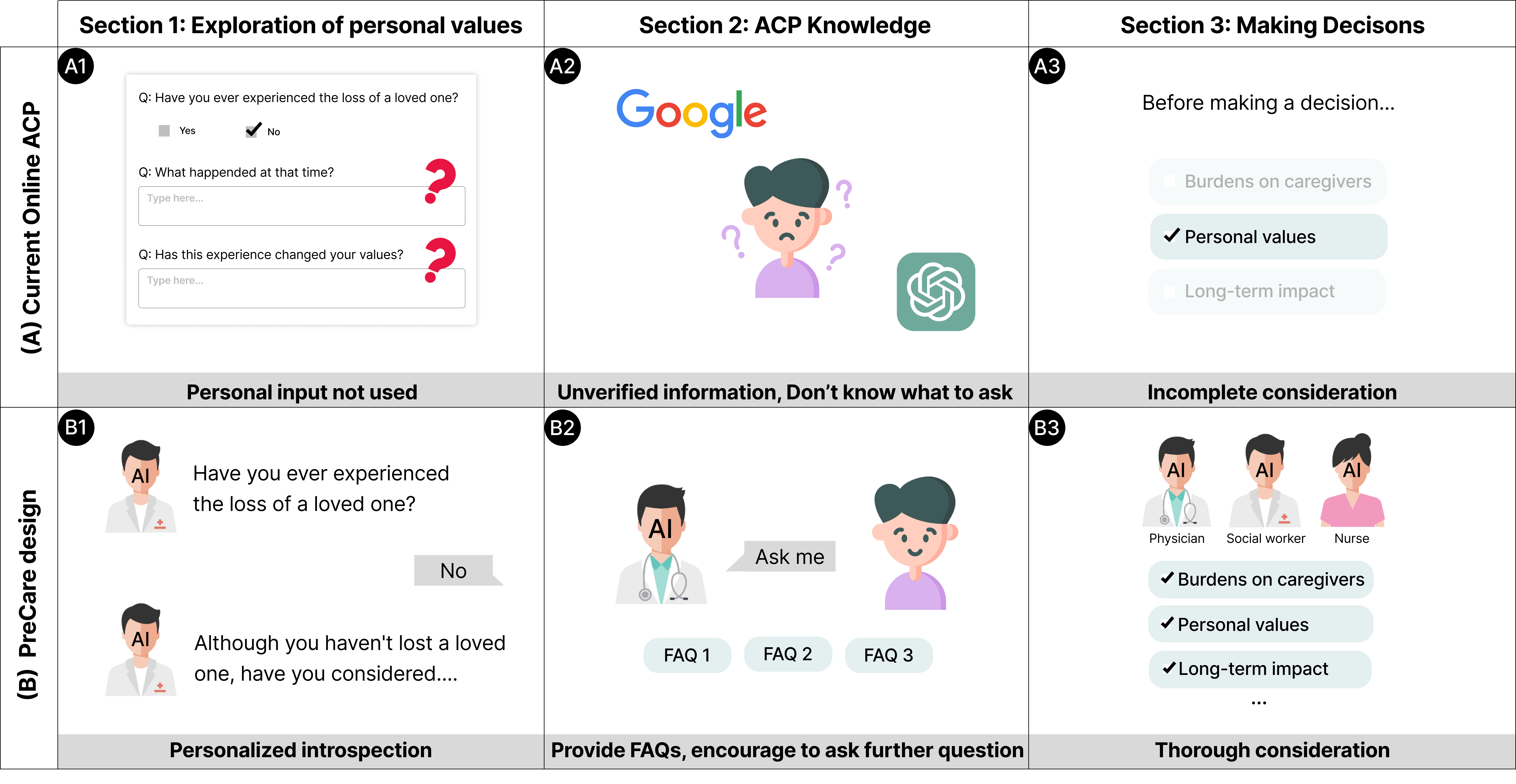}
    \captionof{figure}{    
    Comparison between (1) Current online ACP and (2) PreCare Design in terms of (A) Exploration of Personal Values, (B) ACP Knowledge, and (C) Making decisions. In the Current online ACP: (A1) Participants use a text area to answer questions for exploration of personal values. However, the input is not used for further discussion, and users may simply skip important questions that are crucial for deeper reflection; (A2) If participants have questions, they need to search the internet, but much of the information found online is not verified by professionals. Additionally, some users may be unaware of the knowledge gaps they have or may not know what or how to ask; (A3) Participants miss some aspects before making a decision, and the website does not provide some crucial information before making decisions. In PreCare Design: (B1) We utilized interactive conversations, designed with domain knowledge of social workers,  to facilitate introspection of personal values. Compared to current online ACP platforms, which are limited by a lack of personalized questioning, our assistant asks more follow-up questions to encourage a deeper exploration of personal values; (B2) We integrated physicians' domain knowledge into a prioritized list of frequently asked questions and enabled real-time Q\&A for participants; (B3) By providing comprehensive information and facilitating interactive conversations, with input from physicians, nurses, and social workers, PreCare ensures that participants consider all important aspects before making their decisions.
}
\label{fig:Comparison.png_img}
\end{center}%
}]

 \section{Introduction}

Imagine being unable to make decisions about your own life or death. Research shows that approximately 42.5\% of individuals require decisions to be made about life-sustaining medical treatment in the final days of their lives, yet 70.3\% of them lack the capacity to decide for themselves due to injury or terminal illnesses, such as coma, cancer, Alzheimer’s disease and other forms of dementia~\cite{Silveira2010Advancedirectives}.

While one may think that their surrogates - typically spouses, parents, sons, and daughters~\cite{Torke2014Scopeandoutcomesofsurrogate} - will make the same decisions as them, research has shown that patient–surrogate agreement on life-sustaining decisions has been poor, with up to 

46.6\% disagreement even for spouses~\cite{Spalding2020Accuracyinsurrogateendoflifemedicaldecisionmaking, Uhlmann1988Physiciansandspousespredictions}.

To address this, Advance Care Planning (ACP) was developed to allow individuals to specify their end-of-life, life-sustaining medical care in advance,
to align with their values and goals~\cite{Roberts2024Apersonalizedandinteractive, Silveira2010Advancedirectives}. 
For example, while some may prefer living for as long as possible regardless of the quality of life, others may prefer not to receive life-sustaining medical treatment if the quality of artificially-sustained life would be significantly degraded for themselves and their family.

During the ACP process, individuals explore their personal values, including goals regarding care and quality of life, and record their preferences for potential future medical treatment by completing documentation known as Advance Decision (AD), also known as Advance Directive or a living will. They record their chosen life-sustaining treatment or designate a surrogate decision-maker~\cite{Roberts2024Apersonalizedandinteractive, Silveira2010Advancedirectives}.  

With ACP and AD, patients experience a better quality of life~\cite{Wright2008Associations}, receive less unwanted medical care~\cite{Bischoff2013Advancecareplanning,Detering2010Theimpactofadvancecareplanning,Levoy2023Dontthrowthebaby} and incur lower healthcare costs~\cite{Bond2018Advancecareplanning,Patel2018Effectofalay}. Their families and caregivers also experience reduced anxiety~\cite{BarSela2022Advancecareplanning}.

Advance Care Planning can be conducted through various methods, including: 1) paper-based AD with professional(s) involvement~\cite{BenMoshe2023Advancecareplanning}, 2) web-based AD with remote professional(s) involvement~\cite{Zive2016Implementationofanovelelectronic, legalzoomLivingWill}, 3) Self-guided paper-based AD~\cite{aarpFreeAdvancedirectives, Sudore2007Anadvancedirectiveredesigned}, and 4) Self-guided web-based AD~\cite{Sudore2017Effectoftheprepare}. 

With the involvement of ACP professionals, whether through a team consultation or an individual physician, lawyer, or other medical professional, ACP consultation fosters deeper engagement and exploration of personal values and provides immediate clarifications on complex medical treatments and their consequences~\cite{Sudore2010Redefining}. It also helps individuals consider the quality of life and financial impact of end-of-life decisions on themselves as well as their family members and caregivers~\cite{Sudore2017Definingadvancecareplanning} before making decisions. 
 % Professionals providing ACP consultation can range from a single physician to a specialized ACP team consisting of physicians, nurses and social workers. 
 However, consultation is typically costly with limited availability, especially for full ACP teams~\cite{vanderSmissen2020Thefeasibilityandeffectiveness}, and the quality of consultation can vary widely as
% ACP process takes time, and supporting patients in this process is costly. 
 some physicians report insufficient training or time to conduct ACP discussions during busy clinic visits~\cite{Blackwood2019Barriertoadvancecareplanning, Hemsley2019Anintegrativereview, Genewick2017Characteristics}.
  
Recent development of self-guided ACP websites and apps, such as MyDirectives~\cite{Holland2017Nurseledpatientcentered, Fine2016Earlyexperience}, NVLivingWill~\cite{Klugman2012Anevaluationof2online}, Plan Your Life Span~\cite{Lindquist2017Planyourliffespan, RamirezZohfeld2020Longitudinalfollowup}, PREPARE~\cite{Sudore2014Anovelwebsite, Ouchi2017Preparingolderadultswithseriousillness, Cresswell2017Evaluationofanadvancecareplanning, Sudore2017Effectoftheprepare, Sudore2018Engagingdiverse, Zapata2018Feasibilityofavideobased, Lum2018Improvingafullrange, Freytag2020Empoweringolderadults, Howard2020Effectofaninteractivewebsite} and Koda Health~\cite{Roberts2024Apersonalizedandinteractive} have made ACP widely available and easily accessible compared to clinical consultations. 
  These platforms have been designed for ease of comprehension and usability on web and mobile devices, using videos to augment text and images to explain ACP concepts~\cite{Dupont2022Publiclyavailableinteractiveweb, Sudore2014Anovelwebsite, vanderSmissen2020Thefeasibilityandeffectiveness, McDarby2020Mobileapplicationsforadvancecareplanning}, which enables individuals to complete self-guided forms on their own anytime, anywhere.

While online ACP platforms offer high accessibility, they currently present information and questionnaires without leveraging user responses to ensure that users fully explore their personal values, acquire adequate ACP knowledge, and understand the consequences of their choices. Consequently, users may document decisions that do not properly align with their values.

Recent advancements in AI has made it possible to engage users in a process that mirrors discussions with humans in medical contexts, offering both conversational interaction and medical knowledge.~\cite{Kocielnik2019Harborbot, Thirunavukarasu2023Largelanguagemodelsinmedicine, li2024beyondthewaitingroom}. 

In this research, we explore designing AI technology to bridge the gap between online ACP and clinical ACP consultation.

We first conducted a formative study involving three ACP team consultations at the top-ranked hospital in our region, which included physicians, nurses, and social workers, to deepen our understanding of how medical professionals discuss ACP with their patients. Through this process, we identified three important goals of ACP consultations and how ACP professionals accomplish them.
% \emph{eliciting patient values and medical preferences}, \emph{ensuring knowledge absorption}, and \emph{providing an overview before decision-making}. 
We then conducted a second formative study to identify the specific challenges users may encounter while using self-guided ACP websites in terms of supporting the three goals.

Based on the two formative studies, we developed PreCare, an integrated system featuring three AI assistants to effectively support users in Advance Care Planning (ACP):

\begin{enumerate}
    \item \textbf{Semi-structured interview for introspection of personal values}: As shown in Figure~\ref{fig:Comparison.png_img}(a1), current online ACP provides text fields for users to answer questions about personal values, but their responses are simply unused. Modeled after how clinical consultations interactively converse with users to ensure more complete introspection of personal values, we developed an LLM-based chatbot that conducts semi-structured interviews, as shown in Figure~\ref{fig:Comparison.png_img}(b1).
    
    \item \textbf{Real-time Q\&A with curated and prioritized frequently asked questions (FAQs)}: online ACP users currently have to use search engines (e.g., Google) and general-purpose LLMs (e.g., ChatGPT) , which may deliver information of uncertain accuracy , as shown in Figure~\ref{fig:Comparison.png_img}(b1).

    To ensure correctness, we created a curated ACP knowledge database validated by three experienced ACP physicians. Leveraging Retrieval-Augmented Generation (RAG), we developed an AI assistant to provide accurate, evidence-based answers and highlights the top frequently asked questions (FAQs) ranked by ACP professionals, assisting users who are unsure what questions to ask.

    \item \textbf{Interactive review through personalized impact analysis}: As shown in Figure~\ref{fig:Comparison.png_img}(c1), current online systems do not know whether users have considered all aspects of consequences before making decisions.
    Designed based on how ACP consultations would comprehensively review key aspects that need to be considered before making decisions, we developed an LLM-based chatbot that interactively reviews them with users. The specific aspects have been validated by 3 ACP physicians and 2 social workers, and include the following: "benefits and side effects", "quality of life", "medical expenses", "real-life stories", "caregivers' responsibilities" and "long-term impact" of each of the options of a decision, as shown in Figure~\ref{fig:Comparison.png_img}(c2).

\end{enumerate}

To evaluate the usability and integration of three AI assisants of PreCare, we conducted a usability study (n=12). Results showed that PreCare was rated excellent on the System Usability Scale (SUS).

Furthermore, to better understand the user experience of PreCare's AI assistants, we conducted a user experience study (n=12) and compared PreCare \textit{with} vs. \textit{without} AI assistants. Study results showed that PreCare \textit{with} AI assistants significantly increased knowledge (p<.01), exploration of personal values (p<.01), decisional confidence (p<.05), and was preferred by 92\% of participants.  Note that all studies were approved by the research ethics committee at our university.

In summary, our key contributions include the following:
\begin{itemize}
    \item \textbf{Empirical exploration.} The first formative study provides empirical insights into ACP professionals' domain knowledge for conducting successful ACP discussions and highlights three key goals of consultation. The second formative study reveals that while ACP websites attempt to achieve these goals, users may face challenges due to limited interactivity.
    Through these analyses, we identified critical gaps between clinical consultations and online settings.
    
    \item \textbf{PreCare.} A novel approach of using AI assistants to combine the key benefits of clinical consultation with the accessibility of online ACP, informed by the domain knowledge of ACP professionals, including physicians, nurses, and social workers. PreCare was iteratively designed and implemented in collaboration with ACP professionals, HCI experts, UI designers, and end-users to ensure correctness and usability. PreCare offers a personalized user experience tailored to each individual's values and knowledge level, enhancing the decision-making process by fostering deeper reflection, providing relevant knowledge, and ensuring that decisions are aligned with personal preferences.
    % Iteratively designed and implemented AI assistants in collaboration with ACP professionals throughout the development and evaluation process, ensuring correctness, safety, and relevance through co-design and knowledge validation.

    \item \textbf{Validation.} Evaluation through 2 user studies demonstrating excellent system usability as well as the effectiveness of AI assistants to significantly improve exploration of personal values, ACP knowledge, and decisional confidence.

    \item \textbf{Design recommendations.} Based on the study results, we propose design recommendations for related fields using AI technology. These recommendations aim to guide designers and practitioners in leveraging our insights and building upon our progress to create more effective and user-centered systems.

\end{itemize}

\section{Related Work}

\subsection{Technology for Advance Care Planning}
Several websites and apps have been developed to make advance care planning (ACP) more accessible to the public, extending beyond traditional in-person clinical consultations~\cite{vanderSmissen2020Thefeasibilityandeffectiveness, McDarby2020Mobileapplicationsforadvancecareplanning}. These platforms provide interactive navigation and web-based forms, offering a more engaging experience compared to paper-based pamphlets and forms~\cite{Sudore2017Effectoftheprepare, Schopfer2022Amobileappforadvancecareplanning, Schpfer2023Effectofanapp}.

MyDirectives~\footnote{MyDirectives: \url{https://www.mydirectives.com/}} uses digital forms with multiple-choice and open-ended questions to help users easily document their preferences. Besides digital forms, PREPARE~\footnote{PREPARE:\url{https://prepareforyourcare.org/}} has reconceptualized ACP as a multi-step process, incorporating videos at each stage to engage users~\cite{Sudore2014Anovelwebsite, Sudore2017Effectoftheprepare}. Similarly, the Koda digital ACP platform offers video-based educational content on common life-support treatments and decision-making guides~\cite{Roberts2024Apersonalizedandinteractive}.

While these programs have been widely used, and demonstrated increased documentation rates and improved readiness for ACP~\cite{Roberts2024Apersonalizedandinteractive, Sudore2017Effectoftheprepare, Holland2017Nurseledpatientcentered}, their interactivity remains limited to: 1) navigation, where content is divided into multiple pages for easier consumption, 2) text fields for users to input their thoughts, 3) video stories at each step to guide users, 4) hyperlinks to related web pages, and 5) providing relevant information based on user selections~\cite{Dupont2022Publiclyavailableinteractiveweb}. These available platforms do not support key aspects of clinical consultations.

For example, users' inputs are often unused; users can provide irrelevant or random answers and still complete the documentation. The system also can not offer feedback or guidance tailored to users' inputs.
Additionally, if users have questions, they need to search across multiple website without assurance of accuracy. 
This lack of exploration and incorrect medical understanding may lead to poor or misguided decision-making.

To address this, we propose a novel approach using AI assistants to combine the key benefits of clinical consultations with the accessibility of online ACP platforms. We support interactive conversations where users' input is used to enhance reflection, gain knowledge, and ensure thorough consideration throughout the entire Advance Care Planning process—three critical aspects identified by ACP professionals for effective Advance Care Planning.

\subsection{AI for decision making}
Decision-making can be understood as the conscious process of evaluating various options and selecting the most appropriate one to achieve specific goals~\cite{Morelli2021Decisionmaking}. 
% People gather more information for the best alternative under consideration, and make a better choices~\cite{Lelis2011Informingdecisions}
There has been growing discussion around human-AI decision-making in various fields, such as coordinating teams to control aerial vehicles for photographing ground targets~\cite{Demir2016Teamcommunication},  using AI-assisted group decisions to assess the likelihood of defendants re-offending within two years of their most recent charge~\cite{Chiang2024EnhancingAIassistedgroupdecision}, human-AI collaborative navigation system to facilitate pathologists workflows and accurate scan diagnosis~\cite{Gu2023Augmentingpathologists}, and AI-assisted decision-making system for mechanical engineering designers~\cite{Chong2024Humandesigners}.
Typically, such AI provides suggestions, while humans retain the ultimate authority to make the final decisions~\cite{Buccinca2021Totrustortothink,Wang2022Willyouaccepttheairecommendation,Vereschak2024Trustinai}.

However, prior studies have primarily focused on decision-making processes involving experts or participants with existing background knowledge, who already possess the capability to make decisions but use AI to enhance performance. This highlights a need to further explore the integration of knowledge~\cite{Lelis2011Informingdecisions} and personal value~\cite{Kosaar2023Theimportanceofreflectivepractices}, which are essential for supporting non-experts or the general population, into making informed decisions. To address this gap, PreCare extends beyond decision-making of end-of-life issues and AI-generated recommendations. For advance decision, people may not naturally reflect on their end-of-life values or be familiar with medical knowledge beforehand, therefore, PreCare emphasizes the exploration of personal values and the acquisition of critical knowledge before decisions are made. 
The system integrates three AI assistants that work cohesively to support users make their own decision.

\section{Formative Study \#1: Shadowing of Clinical Consultation}

% The first formative study aimed to identify 

% the pain points associated with self-guided use of ACP (Advance Care Planning) websites. To address these challenges, we conducted shadowing sessions during clinical consultations and interviewed ACP professionals to gain valuable insights. These findings informed the design of a self-guided ACP website tailored for adults.
% !!!!!!!need to say why adults?

% In order to combine the key benefits of clinical consultations with the accessibility of online ACP platforms, we first conducted a series of formative studies, including user experience evaluations with an ACP website developed based on state-of-the-art online ACP systems. And then we conducted shadowing clinical consultations, and interviews with ACP professionals to gain insights on effectively combining the strengths of clinical consultations with online ACP platforms.

The first of our two formative studies aimed to investigate the consultation methods, techniques, and best practices employed by professional ACP teams.

% We conducted a formative study with three ACP teams to deepen our understanding of the critical interactions and key details that require attention during the ACP process.

\subsection{Study Design, Procedure, and Participants}
The research method we used was shadowing followed by semi-structured interviews. We shadowed clinical consultations with three ACP teams from one medical center, ranked as the top hospital in Taiwan
% our region\footnote{Anonymized for review}
, to understand how medical professionals discuss ACP with their patients (in this study, we refer \emph{"patient"} for those who attend advance care planning clinic). The medical center is affiliated with our university, and its director of Family Medicine facilitated the recruitment of three ACP teams for the study. Each ACP team comprised a physician, a social worker, and a nurse, ensuring a multidisciplinary perspective in the consultations.

In Taiwan
% our region
, ACP consultations are conducted in groups, meaning patients typically attend with their family members or friends. One consultation lasted about one hour. We observed five group consultations, with varying numbers of patients signing Advance Decision (5, 4, 2, 3, and 4). Two consultations were shadowed for the first ACP team, one for the second team, and two for the third team. 

At the start of each consultation, the ACP physician introduced the first author—who holds a medical license—as a shadowing researcher and obtained oral consent from the patients. The researcher observed the interactions between the ACP teams and patients without intervening in the consultations. As no patient data was used, and only the interactive dynamics during conversations were documented, no demographic data of the patients was recorded. Following the consultations, the researcher conducted semi-structured interviews , which were recorded, with the three ACP physicians(referred to as PH1~PH3), two social workers(referred to as S1~S2) and one nurse (N1) from ACP teams. The interviews focused on the critical details that require special attention during consultations and the reasoning behind these practices. All ACP professionals had 4 to 5 years of clinical experience in providing ACP consultation\footnote{In Taiwan, the legislation for Advance Decision was passed in 2019.}. The three physicians were Family Medicine specialists with clinical experience of 16, 20, and 23 years, each conducting ACP consultations for 10 patients per month. The two social workers and one nurse all work full-time on patient communication and ACP reservations. They participate in the ACP consultations of 40 patients per month. 

We analyzed the transcribed interview recording using the following approach~\cite{Gu2023Augmentingpathologists}: 
% We analyzed the transcribed interview recording using thematic analysis~\cite{Braun2019Reflectingonreflexivethematicanalysis} to systematically identify and interpret patterns (themes) in the data. 
the second and third authors read the transcripts and coded the data independently, and then the first author reviewed and addressed the disagreements. The findings were subsequently shared with the interviewed ACP professionals for their feedback and to ensure accuracy and corrections where necessary.
The length of each consultation shadowing was about 60 minutes, and the semi-structured interview was about 20 minutes.

This study was approved by the research ethics committee of our university, and was conducted in the ACP clinic at the medical center affiliated with the university.

% We recruited 4 physicians (PH1-PH4), 3 social workers (S1-S3), and 1 nurse (N1) from ACP teams at two medical centers to explore how these insights could benefit current ACP programs. 

\subsection{Findings}\hfill

We identified three key goals of ACP consultations and how ACP professionals accomplish them.
%上面老師改過

% key aspects of what ACP team professionals actually do during consultations and, based on their experience, which areas are particularly important to clarify and address.
%既有的流程中，identified 三個ACP中的應該做到的目標

% \paragraph{\textbf{Outline of ACP consultation}}
% \begin{itemize}
%     \item \textbf{Introduction:} 
    
% \end{itemize}

% \paragraph{\textbf{Aspects emphasized by ACP professionals}}
\begin{enumerate}
    % \item \texbf{Ongoing conversation to }
    \item \textbf{Exploration of patient values and medical preference through "semi-structured interview"}
    % 讓病人自己了解
    % 讓醫生知道
    
    All ACP teams emphasized semi-structured interviews as one of the initial steps for ACP in order to \pquotes{elicit patients' values.}{PH1-PH3}. Without understanding the patient's values, ACP teams can not assess whether the patients' end-of-life medical preferences (advance decision) \pquotes{align with their personal values}{PH1}. As a result, \pquotes{patients are unsure which decision suits them}{PH3, N1, S1}, and the physician \pquotes{cannot provide effective guidance}{PH2}.

    For example, \pquotes{there was a case where someone said nothing during the consultation and ultimately chose to receive all treatments. The accompanying family members were shocked. After the consultation, they kept asking questions and discovered the person actually wished to die with dignity, leading to a swift change in the decision and the need to re-sign the advance decision}{S2}.
    
    However, professionals mentioned most of patients \pquotes{haven't considered their own perpective on end-of-life before}{PH1-2, S2}. Therefore, ACP professionals 
    % will use semi-structured interviews to guide participants step by step . ACP teams
    would ask a pre-determined set of open questions and explore further if needed to help articulate what truly matters to patients.
    For example, at the beginning of the consultation, ACP professionals often start by asking patients, \emph{"what motivated you to consider signing an AD?"} If the patient provides a brief or reluctant response, the professionals may \pquotes{encourage accompanying family members or friends to share their thoughts first, fostering an environment where the patient feels more comfortable sharing personal insights}{PH3}. Alternatively, they may use follow-up questions such as, \emph{"Have you encountered related knowledge or experiences before? If so, in what context or situation? What are your thoughts on these experiences?"} to guide the patient step by step in expressing their inner thoughts.

    In addition, some patients may jump into signing an advance decision without fully understanding what they want due to \pquotes{recent loss of family members or friends}{S1}. When working with this bereaved, professionals engage in discussions \pquotes{about patients' past experiences to help them understand the origins of their thoughts and how these experiences have shaped their personal values, ensuring they not make hasty decisions}{S1}.
    
    This highlights the importance of exploring patient values and preferences as the foundational step in the ACP process, ensuring that subsequent decisions align with their core beliefs and personal needs. 
    % \pquotes{some patients, saddened by the loss of family members, may want to quickly sign an advance decision without truly understanding what they want. Therefore, guiding them to discuss these past experiences to help them understand their values regarding end-of-life care is essential.}{S1} 
    
    % Therefore, it requires ongoing communication and in-depth discussions about what truly matters to the participant (PH1, N1, S1). 
    % For some patients, 可能因為家人過世很傷心，因此會想要趕快簽屬advance decision，但其實並不瞭解自己想要什麼。因此社工師引導他說出這些過往的經驗，以讓他了解自己看待生命末期的價值 (S1)
    % 每個人重視的價值不一樣，但是他們不一定了解或想過自己看待生命的方式。因此需要藉由(PH2)
    % one physician mentioned that \emph{"people don’t always realize whether they’ve explored their personal values regarding end-of-life issues enough; it requires ongoing communication and in-depth discussions about what truly matters to them."}
    % However, when 

    \item \textbf{Provide ACP knowledge by walking through ACP topics and answering questions} 
    % \item  \textbf{Active discussion to address "hesitation to ask" and cognitive bias questions. } 

    In clinical consultations, ACP teams spend significant time explaining crucial information about ACP, allowing participants to ask questions whenever they have concerns. This is because medical treatments and conditions are often \pquotes{unfamiliar to the general population}{N1}, and patients are usually \pquotes{hesitate to ask questions}{S2, PH4}. Misconceptions about illnesses or overly optimistic expectations regarding treatments can lead to \pquotes{the medical care provided contradicts expectations}{PH2, S2}. For instance, \pquotes{many people view a nasogastric tube as a routine medical procedure, but in reality, it is extremely uncomfortable. While choosing to use a nasogastric tube may extend life, it comes with the burden of significant discomfort}{PH3}.
    
    % In clinical consultations, ACP teams spend significant time explaining crucial information about ACP, allowing participants to ask questions whenever they have concerns. However, users may encounter challenges such as: 1) having questions but feeling hesitant to ask (PH1), or 2) being unaware of their own questions due to cognitive biases(S2, PH4). These challenges can limit their understanding or reinforce misconceptions about crucial medical knowledge needed for decision-making.
    
    % To address this, the ACP team would explain essential knowledge by highlighting common misconceptions held by the public. They then ask the patient, \emph{"Is there anything unclear that needs further explanation?"} to ensure comprehension and address any concerns.
    
    To address this, ACP teams break down essential medical knowledge for ACP into smaller sections, \pquotes{explaining one part at a time and asking if participants have any questions before proceeding}{PH2}. Directly asking whether they have questions about the entire AD or all medical treatments may not yield responses if they are unsure how to articulate their concerns. Thus, providing some information and then pausing to check for understanding after each section is crucial before advancing to the next stage.
    It was also noted that \pquotes{when relatives or friends participate in ACP together, if someone asks the first question, it tends to encourage others to feel more comfortable asking their own questions}{PH3}. 
    
    % ACP teams encourage participants to voice their concerns at any time. First
    
    % For example, directly asking whether they have any issues with the entire AD might not elicit a response if they are unsure how to articulate their questions. Therefore, after explaining each section, \pquotes{I would pause to ask if they have any questions, actively encourage inquiries, and ensure their understanding before proceeding to the next stage.}{PH2} 
    
    % For the second challenge, medical professionals ask them questions in return to ensure that they truly understand the applicable scenarios and decisions that can be made with an Advance Decision (S1, N1, PH1). 
    
    In short, medical professionals emphasized that the ACP process involves more than merely conveying knowledge in a one-way manner; it is also interactive, designed to draw out participants' underlying uncertainties and ensure that they have no lingering questions. This highlights the importance of ensuring knowledge acquisition, as patients need to fully understand critical information to make informed and realistic decisions about their care preferences.
    % In clinical consultation, ACP teams 會花很多時間解釋有關於ACP的重要知識，並讓participant有問題時可以詢問。然而使用者還是會有: 1) they have questions but are hesitant to ask (PH1), or 2) they are unaware that they have questions(S2, PH4). For the first situation, one solution is to encourage them to voice their concerns. One also noted \pquotes{When relatives and friends participate in ACP together, if someone asks the first question, it tends to encourage others to feel more comfortable asking their own questions}{PH3}; for the second situation, medical professionals would 反問他們一些問題，確認他們是否真的了解Advance decision的適用情境與可以決定的事項。
    
    % medical professionals mentioned that it’s important to encourage them to voice their concerns, while for the latter, continuous interaction is needed to prompt them to express their understanding, ensuring that any uncertainties are addressed.

    %因為病人不一定知道自己不知道什麼，因此需要去proactively asked (S2, PH1)
    %
    
    %However, in the current SOTA website, the knowledge of ACP is limited to images, text, and videos. The participants屬於被動吸收知識(P5)。if participants have any questions, \pquotes{I could only search online, but I was not sure if the information was accurate.}{P2, P14}. In addition, knowledge needed to make informed decisions is lacking, including \pquotes{details of medical conditions, life-sustaining treatment, and artificial nutrition, such as prognosis and impact}{P1-2, P7-8, P10-11, P14}. 

    % 透過分向檢察來看自己決定的取捨
    % 利用細項檢視得方式，來綜觀醫療決定的取與捨
    \item  \textbf{Review the implications of decisions through personalized impact analysis} 

    Signing AD is one of the final steps of the ACP process. However, some patients may \pquotes{overlook the consequence behind the decision}{PH3}. For example, S1 noted \emph{"some people from financially struggling backgrounds choose to accept all treatments, but they may overlook scenarios where a vegetative state requires long-term care and expenses."}
    
    Therefore, medical professionals would guide patients to consider multiple aspects before finalizing their choice, encouraging \pquotes{both emotional reflection and rational thinking}{PH2}. Throughout the discussion, ACP teams would prompt the patients to consider: 1) whether the quality of life resulting from their decision is what they would desire, 2) the potential medical expenses, 3) the long-term impact of their decisions, 4) the benefits or side effects of each option, and 5) whether the options reflect their personal values and knowledge, ultimately enabling them to make informed decisions. 
    % For example, \pquotes{People may ignor financial burdens for vegatative state}{}
    %大家可能會忽略什麼? 

    During the signing process, the ACP team also recognizes that content complexity may lead to patient misunderstandings. To address this, physicians, nurses, and social workers collaboratively assist patients in completing each section. For each clause, the team reviews what is being signed, and then encourages the patients to \pquotes{verbalize their choices}{PH2, S2}, and ensures that they fully understand the implications of their decisions. If the patient hesitates between two options, the ACP team also helps \pquotes{compare them based on the aforementioned aspects.}{PH1-3}
    
    % This interactive process allows users to ask questions in real-time and provides immediate access to professional guidance.

    This approach highlights the importance of analyzing impact of decisions before finalizing such significant decisions.
    % During signing, 因為內容很複雜，如果只是patient自己填可能會不太了解但仍繼續填下去。因此ACP team會讓使用者在physician, nurse, social worker的幫助下一一填寫，由acp team說出目前正在簽屬什麼條目，並讓使用者填寫時說出自己的選則。這樣的方法能讓使用者知道自己有問題隨時可以問身邊幫他一起簽屬的專家
    
    % This highlights the importance of careful consideration before making a decision.
%medical expenses, benefits or side effects, long-term impact, real-life stories, burden on caregivers, personal values, knowledge of ACP
    
    % However, to the SOTA website, decisions including "try all treatments," "do a trial," "reject all treatments," or "delegate decisions to my proxy" have less mention on \pquotes{impact on both me and my family}{P6-7, P10, P14}. As \pquotes{carefully discuss and consider all aspects, without overlooking any details, before making a decision.}{P3-8} is important to ensure the decision reflect on personal values, knowledge of ACP, and well-thinkg of each decision.
\end{enumerate}

\section{Formative Study \#2: User Experience of Online ACP}

The second formative study aimed to assess the current user experience with self-guided Advance Care Planning (ACP) websites. It focused on three key goals identified in the first formative study: 1) How does the ACP website help users explore personal values? 2) How does it assist in acquiring medical knowledge related to ACP? and 3) How does it facilitate making informed decisions?

\subsection{Design of Online ACP Website}
We designed an ACP website that combined the key features of the following three state-of-the-art online ACP and the most authoritative ACP guide in our region to ensure compliance with local laws and appropriate context:

\begin{itemize}
    \item  \textbf{MyDirectives~\footnote{MyDirectives: \url{https://www.mydirectives.com/}}}: MyDirectives was one of the first programs to emphasize combining the elements of patient values with reflection on treatment preferences for online ACP approaches~\cite{Fine2016Earlyexperience}.
    % It organizes its content into four key topics: documenting personal values, specifying medical treatment preferences, addressing final days, and designating Healthcare Agents.

    \item  \textbf{PREPARE~\footnote{PREPARE: \url{https://prepareforyourcare.org/}}}: 
    PREPARE reconceptualized ACP as a multi-step process to engage participants, and used videos at each step to better illustrate the concepts~\cite{Sudore2014Anovelwebsite}. 
  
    \item  \textbf{Koda~\footnote{Koda Health was not available to the public at the time of this research. Therefore, we identified its key design benefits from its publications and website description.}}: Koda incorporates and provides the most comprehensive information on common life support treatments used in end-of-life situations~\cite{Roberts2024Apersonalizedandinteractive}. 

    \item  \textbf{Localized ACP guide\footnote{The guide was specifically designed for our region <anonymized for review>, with permission for use in our system and studies.}}: This guide was developed by a public hospital in our region and followed the easy-to-read guidelines~\cite{Milan2021Informationfroall}. We used the images, questions, and information related to ACP and AD from the guide to illustrate the concepts, ensuring the content is comprehensive and easy to read. 
\end{itemize}

The scope of ACP decisions varies by country, and our design aligns with the regulations of our country, including five medical conditions~\footnote{The five medical conditions include: terminal stage of illness, an irreversible coma, a permanent vegetative state, severe dementia, and other disease determined by the government.} which may impair consciousness in making a decisions, as well as five types of life-sustaining treatment~\footnote{Life-sustaining treatment refers to necessary medical measures that can prolong the lives of patients, including cardiopulmonary resuscitation, invasive mechanical ventilator, extracorporeal membrane Oxygenationblood products, blood products, and antibiotics against severe infections.} and two types of artificial nutrition and hydration treatments\footnote{Artificial nutrition and hydration refers to the provision of food or fluids via tubes or other invasive means, including nasogastric tube and gastrostomy.} in the website to provide knowledge regarding treatment before the user makes their decision.

Based on the localized ACP guide, our ACP website includes three sections: 1) What matters most to me, including 6 topics of discussion; 2) What is Advance Care Planning; 3) Making Advance Decision. Table~\ref{tab:topics_in_precare} shows the details of the website. 

Two researchers designed the ACP website using the materials described above. Inspired by MyDirectives, we ensured the inclusion of multiple questions to explore personal values toward end-of-life of participants. Drawing from PREPARE's design, each section features several corresponding videos to engage participants\footnote{The videos were provided by a non-profit foundation in our region and were used with permission.}. Additionally, inspired by Koda, our website incorporates the aforementioned scope of ACP coverage and ensures the information aligns with local laws.

In addition, the website includes important interactive functionalities~\cite{Dupont2022Publiclyavailableinteractiveweb}, such as a text area for answering questions, a text-to-speech option, a save button, a progress display, hyperlinks to other web pages, and pre-determined content and feedback tailored to the user's input~\footnote{For example, pressing a button would cause the information to pop up.}. 

Subsequently, an ACP physician and two social workers from shadowing study went through the entire process of the website, providing feedback and addressing any unclear statements to refine the website further.

% then reviewed to ensure its relevance and completeness based on previous research~\cite{Dupont2023Definingthecontentofawebsite, Dupont2022Publiclyavailableinteractiveweb}. 

\subsection{Study Design, Procedure, and Participants}
We recruited a total of 14 participants (P1-P14), comprising 7 males and 7 females, with ages ranging from 19 to 69 (mean=35.6, SD=18.1), via posting on social media. 

Participants were asked to complete a consent form and a demographic survey. Following that, we introduced the study details, encouraging them to use the think-aloud method~\cite{BAXTER2015Understandingyourusers, Schopfer2022Amobileappforadvancecareplanning} while interacting with the website. Specifically, for each interaction with the website or when answering each question, they were encouraged to elaborate on their thoughts. Participants used the website in a quiet setting in our lab. After using the website, we conducted a semi-structured interview focusing on the \emph{"challenges"} and \emph{"potential needs"} of ACP. The entire study lasted about 60 minutes, and participants received nominal compensation.
% equivalent to USD \$10. 
The study was approved by the research ethic committee at our university.

The sessions were fully transcribed. Similar to the previouse study, the second and third authors read the transcripts and coded the data independently, and then the first author reviewed and addressed the disagreements.

\subsection{Findings}

% For three aspects, we identified unmet needs from the current ACP experience:

We summarize how online ACP supports the three key goals from the first formative study, along with the current challenges:
% ACP這三個東西是重要的。我們發現網站也有試圖去解決 1 2的問題，但我們發現他不足以支持。而3是目前沒有提供這個服務(因為是分開的，無法綜合來討論)。

\begin{enumerate}
    \item \textbf{Static questionnaires resulted in inconsistent and limited user engagement.}
% 為了讓使用者想一下進行深度思考，所以設計這個框讓使用者填
% 但只是一次性回答，無法得到反饋，因此回答的還是很淺
% ACP 網頁的目的是希望你去思考一下，先進行深度思考。
%一次性的填答、互動，所以無法更深入的探索。填的好也不會有任何反貴。不足以他有深度探索
% 有受過訓練的人才能自己不斷內省深化
Current ACP websites used static questionnaires and input, such as text areas and checkboxes, to ask participants to reflect on their values before making decisions. 
However, the user responses are not used to in subsequent steps, leading participants to express concerns like \pquotes{afraid of inconsistency}{P1, P4-6, P10-11, P13} and \pquotes{lack motivation to type more.}{P3, P5, P11}

The system can not detect consistency because the data was not used. As P4 stated, \emph{"I typed that what matters most to me is the quality of life rather than its length. However, if I later choose the option for aggressive resuscitation, the system does not highlight this contradiction."} 

In addition,  regardless of what users typed, there is \pquotes{no feedbacks}{ P1, P5-6, P10} or \pquotes{no follow up questions}{P1, P4-5}. Some participants reported that limited engagement made them \pquotes{kept answers simple, leaving some thoughts in my mind}{P11}. P5 emphasized that \emph{"these questions are good to think, but unless someone regularly engages in self-reflection, they can't deepen their thoughts without any feedback. I think most people just fill it (question) out quickly with shallow thinking."} 
The lack of interaction in the system leads to users not wanting to fill it out, and since it doesn't check if they do or not, this may result in insufficient exploration of personal values.

%讓使用者不想思考、紀錄太多
% 類似的keywords要quote出來
% quote短一點
% As mentioned earlier, user inputs are not utilized, making it impossible for the system to determine whether the decisions made align with their values.  Similarly, P10 mentioned, \emph{"There’s no double-checking, so I don't feel confident in making this decision."}
% One reason is that the text areas are static (), and the inputs were treated as independent entries, and were not used to guide or inform subsequent steps. P11 mentioned \emph{"Even if I write a lot, there’s no reward or benefit. After completing one question, I just move to the next—it feels empty. So, I kept it simple, leaving some thoughts in my mind instead of writing them down."} In addition, P5 emphasized that \emph{"these questions are good to think, but unless someone regularly engages in self-reflection, they can't deepen their thoughts during the process. I think most people just fill it out quickly with shallow thinking."}

%Additionally, the Advance Care Planning (ACP) process demands significant time and cognitive effort. The self-guided nature of having to think and fill out everything independently left participants \pquotes{feeling mentally fatigued}{P10, P13}

%心裡會想很多，但會覺得只是在填空格因此

\item  \textbf{Lacks support for user inquiries and clarifications.}
% 被動的知識吸收
% 為了提供ACP必要知識，會有圖片、影片、文字來說明。
% 缺乏沒有check point、不知道使用者有沒有問題。
% 就算有問題也不能發問
Current online ACP use images, text, and videos to help participants understand medical treatments and advance decisions. However, several participants had questions and had to \pquotes{to search on the internet}{P2-3, P12, P14}. The \pquotes{lack of integration}{P14} required users to search for and visit multiple external websites, which interrupted the ACP process and was time consuming. Furthermore, \pquotes{after I Googled for answers, I wasn’t sure if the information was correct.} {P2}

\item  \textbf{Lacks support for personalized analysis of decisions and consequences.}
% ACP 提供的內容是零散的、一個一個的。
% 沒有整合性的評估 
% 使用者沒辦法綜合評估
% 分向零散的討論 
% 有沒有什麼問題沒有人告訴我
%沒有宏觀的角度去看填的內容彼此之間的關聯
Current online ACP do not utilize user input at all, making it hard to provide tailored analysis of decisions to users' needs. 

When filling out the advance decision on the current website, participants reported difficulty in thinking about the \pquotes{consequences of decisions}{P1-2, P7-8, P10-P11} on their own, and \pquotes{lack of confidence}{P6-7, P10, P14} due to fear of missing details.

For example, P1 noted \emph{"I care about the decision's impact on my family as a single-parent household, but there is no information tailored to such family situations like mine."}  P14 also mentioned, \emph{after making a decision, I'm not sure if I fully understand it. I'm afraid there might be aspects I've missed.}
\end{enumerate}
\begin{table*}[]
\centering
\caption{Topics covered in the ACP website, which was based on several state-of-the-art programs. The content was reviewed by a physician and two social workers to ensure its relevance and completeness. }
\label{tab:topics_in_precare}
\begin{tabular}{ll}
\hline
Section                                                & Detail                                               \\ \hline
\multirow{6}{*}{1. What Matters Most to Me}        & Topic 1: My Likes and Joys                           \\
                                                       & Topic 2: Do I want to know what illness I have? Why? \\
                                                       & Topic 3: Past experience of losing loved one         \\
                                                       & Topic 4: My fear when I am ill                       \\
                                                       & Topic 5: My preferred place of death                 \\
                                                       & Topic 6: My preferred burial method                  \\ \hline
\multirow{4}{*}{2. What is Advance Care Planning} & Overview of the Implementation of Advance Decision   \\
                                                       & Why is Advance Decision important?                   \\
                                                       & 5 medical conditions impairing decision making         \\
                                                       & Life-sustaining treatment and artificial nutrition   \\ \hline
\multirow{5}{*}{3. Making Advance Decision}            & Considering a potential proxy decision-maker         \\
                                                       & Thinking before making a decision                      \\
                                                       & Making my own Advance Decision                         \\
                                                       & My proxy decision-maker                              \\
                                                       & Other considerations of my wish                      \\ \hline
\end{tabular}
\end{table*}
\section{Design of PreCare}
% Based on the needs and challenges identified from the formative studies, our insight from clinical consultation is to leverage Large language models to help participants explore personal values, gain knowledge of ACP, and review the consequences of decisions. Three AI assistants are integrated into 3 sections of the ACP website. 
% We used the term \emph{PreCare} to describe the whole service with the AI assistants.
%To better support participants in achieving three key goals during online ACP, 
We drew insights from clinical consultations and explored how recent AI advancements, including Large Language Models (LLMs), Retrieval-Augmented Generation (RAG)~\cite{Lewis2020AdvancesinNeuralInformationProcessingSystems}, and LLM-EVAL~\cite{lin2023llmevalunifiedmultidimensionalautomatic}, can be designed to enhance online ACP. 
The three key design goals are as follows:
%Our aim was to improve the online ACP experience with the following design objectives:
\begin{enumerate}
    \item \textbf{G1: Thorough introspection of personal values.} In line with the ACP team's practice of "semi-structured interviews," the system should engage with users interactively and analyze user responses to ensure sufficiently thorough introspection. 
    %Feedback was given, and follow-up questions
    \item \textbf{G2: Effective ACP knowledge acquisition.} %Expandable FAQ
    To help users acquire ACP related medical knowledge, the system should be capable of answering user questions in real-time with correct information.
    \item \textbf{G3: Personalized review of decision consequences.}
    To help users align their ACP decisions with their personal values, the system should help users review and recap key consequences before they finalize ACP decisions and answer questions that users may have.
    
\end{enumerate}
% We explore how to apply recent advances in AI, including conversational chatbots using Large Language Models (LLMs), Retrieval-Augmented Generation (RAG), and LLM-EVAL, user interface design to enhance online ACP.

One cross-cutting advantage of ACP consultations is the real-time, two-way interaction, enabling clinicians to elicit user responses and address questions immediately.
Recent advances in conversational user interfaces (UIs) demonstrate that LLM-based AI can exhibit conversational capabilities comparable to those of human interactions \cite{li2024beyondthewaitingroom, Albers2024letstalkaboutdeath}. Such chatbots can engage in open-ended dialogue guided by prompt design, provide relevant responses to user input \cite{roller2021recipes}, and dynamically adapt to conversation flow based on previously discussed content \cite{Kumar2023Exploringthepotentialofchatbots}. For each of our three design goals (G1–G3), we developed distinct LLM-based assistants to provide real-time, two-way interaction, each employing different AI technologies optimized to fulfill its respective objective.

%prompt design, iterative design with 
\subsection{Design Process and AI Safety}

We used an iterative approach to refine the AI assistant UI designs and functionality. Three HCI researchers initially designed the assistants based on insights from the two formative studies. These designs were then tested by two end-users (one male and one female, age 26 and 39), four physicians (three from formative study \#1 and one family medicine doctor with over 30 years of experience), and two social workers (both from formative study \#1), then the UI and assistant functionality were iteratively refined. In addition, we consulted a professional UI designer for feedback three times during the design and development process.

\begin{figure*}[h]
    \centering
    \includegraphics[width=\linewidth]{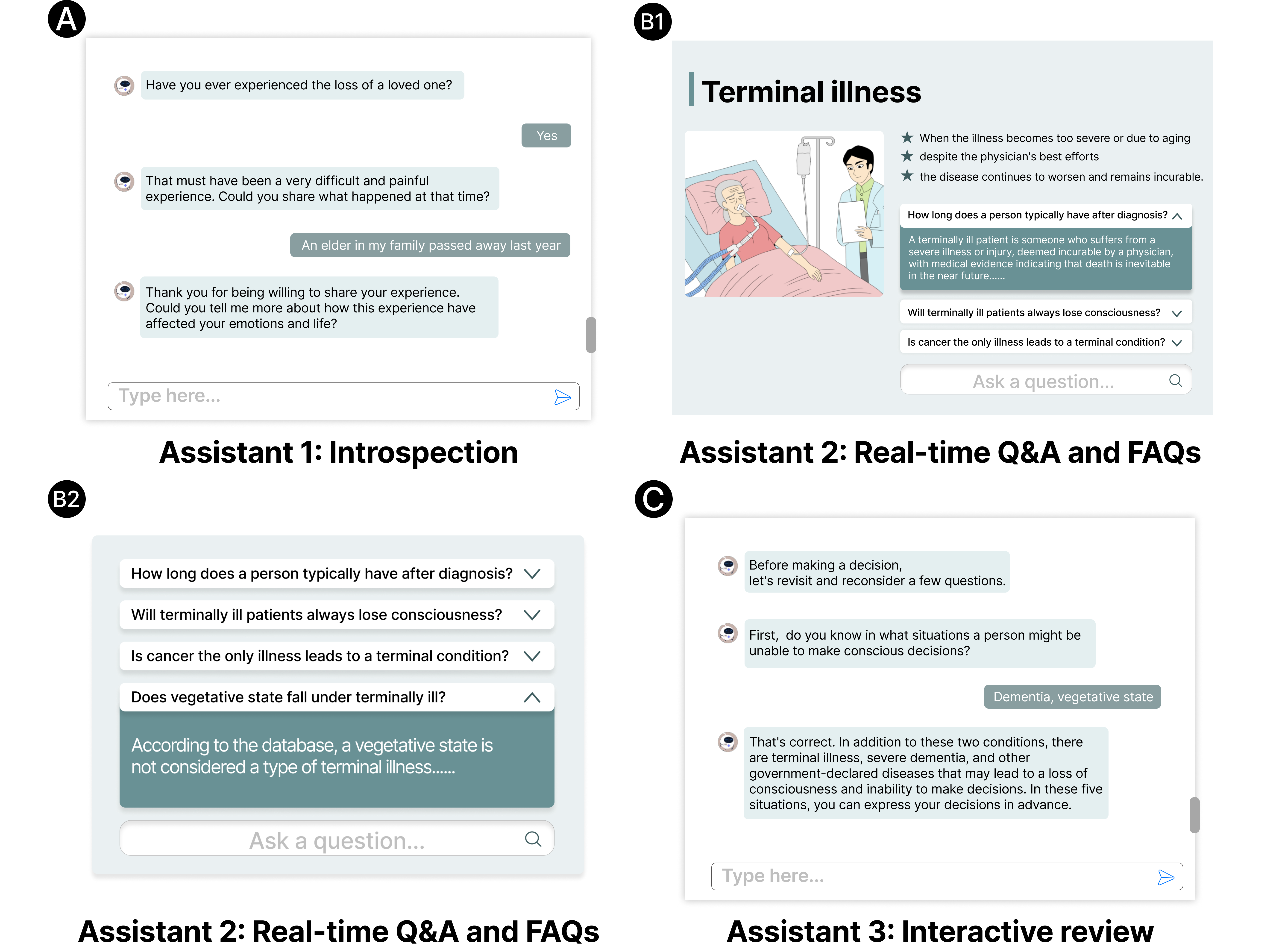}
    \caption{Screenshots of the 3 PreCare AI Assistants: (A) two-way conversation to ensure thorough introspection of users' personal values. (B1) Each key medical knowledge topic includes prioritized, top 3 FAQs and a text field for real-time Q\&A. (B2) Shows an actual Q\&A example from the study (P10), the Q\&A was appended to the FAQ to enable users to continue asking additional questions. 
    %ed a question on the same page as in (B). PreCare responded to the question using the verified database and appended the answer to the bottom of FAQs. 
    %The participant could then ask another question using the placeholder. 
    (C) conversational assistant reviews important aspects before making final decisions. 
    % In-Context FAQ: Features a list of questions within each section. Participants can type their own questions in the same section, where the LLM-based assistant responds based on a database using Retrieval-Augmented Generation (RAG).
    }
    \label{fig:2designs.png_img} 
\end{figure*}

\begin{figure*}
    \centering
    \includegraphics[width=\linewidth]{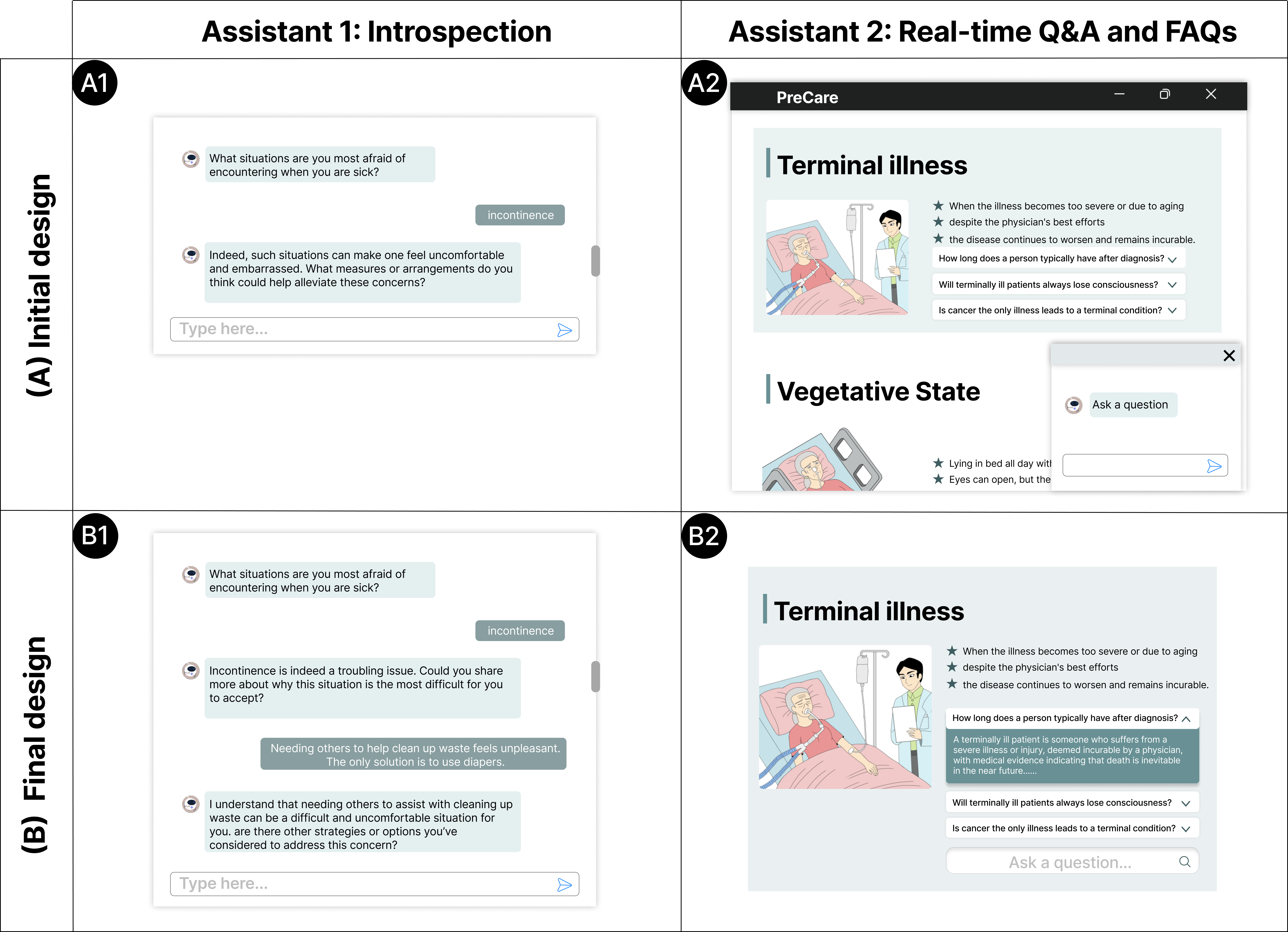}
    \caption{UI and Assistant Design Iteration: (A) Initial vs. (B) Final designs. For Assistant 1, personal value exploration, (A1) testing of the initial design revealed limited follow-up capability; (B1) we added a response evaluation system to enhance follow-up capability. For Assistant 2, ACP knowledge, (A2) the initial design featured a standard chatbot interface for Q\&A; (B2) based on feedback, we integrated Q\&A into the FAQ to improve usability. 
    }
    \label{fig:before_after.png_img} 
\end{figure*}

In this paper, we used the term \emph{PreCare} to describe that website augmented with the 3 AI assistants.

 \subsubsection{AI Safety and Liability}
To prevent bias during the advance care planning process, PreCare implemented several safeguards: 1) The AI assistant provides only feedback and follow-up questions for exploring personal values, encouraging user introspection; 2) Real-time questions are answered using a verified database; 3) Important considerations are presented with complete, balanced information validated by physicians and social workers; 4) Researchers continuously monitored conversations to ensure no identifiable data was recorded, and to verify system safety, accuracy, and absence of bias.

\subsection{Assistant 1: Semi-structured interview for introspection of personal values} \hfill
\paragraph{\textbf{Initial design:}}
To motivate users to share more personal values and think more deeply, we designed the first assistant for personal value exploration of the website. We implemented an LLM-based chatbot to engage in conversation with users, as shown in Figure~\ref{fig:2designs.png_img}(a). 
 Inspired by previous studies on conversational chatbots~\cite{Albers2024letstalkaboutdeath, li2024beyondthewaitingroom} and insights gained through clinical shadowing, particularly from social workers, we designed our prompt in two parts. The first part consists of behavior instructions, while the second part covers basic instructions. 
 
The first part mimics how social workers facilitate the exploration of personal values through "semi-structured interview", including: 1) defining a clear goal for the discussion, 2) avoiding direct sensitive questions when addressing end-of-life topics, 3) incorporating key questions in the system that aid in exploration (pre-determined questions), and 4) emphasizing the importance of feedback and empathy throughout the conversation.
    
The second part instructs the assistant to engage in more fluid discussions with participants. For example, unlike the baseline, where all questions are presented at once, the AI assistant in PreCare asks one question at a time and provides feedback based on the user's input. The prompt also instructs the assistant to skip questions that have already been answered. If the response is vague, the assistant asks follow-up questions to clarify personal values. 

% To prevent the introduction of biased feedback or questions, and to avoid providing recommendations, we iteratively tested the prompt with end-users and social workers. 
The AI assistant in this design was built using the latest version of OpenAI's GPT-4o~\footnote{GPT-4o: \url{https://openai.com/index/hello-gpt-4o/}} at the time of the research (2024 summer). We preferred GPT over LLMs specifically designed for medical tasks because we prioritized conversationality alongside medical expertise, especially given the chatbot's role in facilitating reflection rather than diagnosis~\cite{li2024beyondthewaitingroom}. The model's temperature was set to 0.6 to allow flexibility in providing diverse responses while keeping the discussion focused on the given topic. The prompt details can be found in the Appendix Table~\ref{tab:promptdesign1}.

\paragraph{\textbf{Feedback and improvement:}}

% During design, we found that if the prompt was simply "ask a follow-up question if the user's input is vague," the follow-up ability was weak. 
% 使用者使用發現，
As one end-user commended \emph{"the AI assistant provided feedback but did not generate many follow-up questions tailored to my input."}
%In formative studies, medical professionals assessed whether a patient's response was sufficient for exploring values. 
To address this, we implemented an evaluation system based on LLM-EVAL by Lin et al.~\cite{lin2023llmevalunifiedmultidimensionalautomatic}. The user's input was evaluated based on relevance, appropriateness, and content. The script would reference these evaluation scores each round and prompt follow-up questions in areas where the score was low. The prompt evolved from the previous research prompt of "ask a follow-up question if the user's input is vague"~\cite{li2024beyondthewaitingroom, Albers2024letstalkaboutdeath} to a more structured method. The refined version, as shown in the Appendix Table~\ref{tab:promptdesign1}, includes evaluation criteria and the statement, "You can reference the evaluation scores and ask follow-up questions if the user's response is not satisfactory."
The example conversations of before and after refinement are shown in Figure~\ref{fig:before_after.png_img}.
 \subsection{Assistant 2: Real-time Q\&A with curated and prioritized frequently asked questions (FAQs) to enhance the effectiveness of knowledge acquisition} 
 \paragraph{\textbf{Initial design:}}
 To enhance the effectiveness of ACP knowledge acquisition using the website, we implemented the second AI assistant to support an interactive interface for more efficient learning, encouraging users to actively engage with the content, and seek answers based on domain knowledge. As shown in Figure~\ref{fig:before_after.png_img}(a2), the second AI assistant has 1) an expandable FAQ design corresponding to each key piece of information and 2) one chatbox at the right corner of the screen~\cite{tripaiAI, facebook} to ask additional questions. 
 
 The FAQs are designed with four considerations: 
1) Organized by Sections: FAQs are grouped into sections to provide comprehensive information on each topic, simulating the "walking through ACP topics" practice used by ACP teams.
2) Expandable interface: FAQs are designed to be expandable, enhancing interactivity and making the learning process more engaging for users.
3) Encouraging Questions: To address "hesitance to ask questions," FAQs could be seen as questions posed by others. This design encourages users to feel comfortable asking their own questions, inspired by insights from clinical shadowing.
4) Prioritized Content: FAQs are prioritized to provide either basic or in-depth information based on individual preferences. The most important FAQs are displayed first to streamline access to essential details.

We incorporate FAQ because it helps users gather knowledge efficiently~\cite{nngroup_faqs_value,strategybeam_faq_benefits}. Making an advanced decision requires understanding complex medical knowledge, such as treatment options and the conditions under which one may lose decision-making capacity. FAQs are particularly effective for simplifying this information and addressing common questions. 
 
 The FAQs were derived from formative studies, including but not limited to topics such as the \emph{"discomfort"} and \emph{"success rate of regaining consciousness"} of life-sustaining treatments, what happens to the body during 5 medical conditions and life-sustaining treatment and addressing \emph{"common misconceptions."} Four physicians reviewed them for completeness, accuracy, and importance in Advance Care Planning. As a result, 33 expandable FAQs are included on the section 2 of the website.

% \fixme{walk through, FAQ make each topic more detail know how}
% Additionally, the FAQ design can be viewed as an ongoing interaction between medical professionals and participants, addressing gaps in knowledge that users may be unaware of, holding the potential to address the \emph{"don't know what they don't know"} issue.
    
For asking additional questions, 
% as shown in Figure~\ref{fig:2designs.png_img}(c),
we integrated GPT-4o with Retrieval-Augmented Generation (RAG), using a database from publicly available hospital data (content also verified by 4 physicians). This ensures that the answers provided align with ACP domain knowledge. Additionally, GPT-4o is capable of addressing diverse medical examinations~\cite{Kipp2024fromgpt35togpt4o, Gajjar2024Evaluatingtheperformanceofghatgpt4o}, enabling it to answer general medical questions. 

%The placement of the placeholder mirrors how ACP professionals encourage participants to ask questions after each small section.

\paragraph{\textbf{Feedback and improvement:}}
When testing, the participants reported that the interface required \pquotes{frequent eye movement}{2 end users and 1 UI designer}, which could reduce user motivation to engage with the system. The user needs to see the content and the FAQs on the top, and ask additional questions at the bottom. In addition, as more questions are aksed in the chatbox, it is \pquotes{hard to find previous answers}{2 end users} 
Therefore, we improved the design by introducing \emph{in-context} prioritized FAQs and real-time Q\&A, integrating these functionalities directly alongside the main content. As shown in Figure~\ref{fig:2designs.png_img}(b) and (c) and Figure~\ref{fig:before_after.png_img}(b2), this design minimizes eye movement, enhances user engagement, and ensures immediate access to relevant information, and better simulates ACP teams' practices (the ACP professionals encourage participants to ask questions after each small section),

 \subsection{Assistant 3: Interactive review through personalized impact analysis} 
 
 \paragraph{\textbf{Initial design:}} To facilitate the users' decision tailored to their needs and ensure participants considered all important aspects before making decisions, we designed the third AI assistant in section 3 of the website: 1) provides several key factors that physicians identified as crucial during making decision. These include "benefits or side effects", "quality of life", "medical expenses", "real-life stories," "caregivers' responsibilities," and  "long-term impact" of each option (the implication of the decision); 2) the AI assistant would help participants review through these aspects by proactively discussing and asking whether they had considered each factor, and discuss the difference of each decision, mimicking the process of a physical consultation (impact analysis and ensuring value alignment), as shown in Figure~\ref{fig:2designs.png_img}(d). In addition, the system allows users to inquire about the consequences of any options and compare them based on personal needs at any time, providing recommendations sourced from the database.

 The prompt for the assistant is divided into two parts. The first part, behavior instructions, mimics how ACP teams guide patients to carefully consider all aspects before making a decision. The assistant's main goal is to discuss preferences in the event of losing consciousness due to five specific medical conditions. The assistant begins by reviewing personal values and the participant's knowledge of Advance Care Planning (ACP) to ensure they understand the circumstances under which an Advance Decision would apply. Following this, the assistant thoroughly explores all relevant aspects before making decision.
    %A brief review of personal values, A brief review of knowledge of ACP, Thorough considerations before making decision
The second part was similar to the first assistant. The prompt was also iteratively designed with input from end-users and experts to ensure the prevention of biased feedback or questions. The prompt details can be found in the Appendix Table~\ref{tab:promptdesign3}.

\paragraph{\textbf{Feedback and improvement:}}
After testing, ACP professionals emphasized the importance of AI providing balanced information that does not favor any specific choice, as these decisions have no definitive right or wrong answers within end-of-life preferences. Therefore, all information presented to users, either directly or via the AI assistant, was grounded in validated data reviewed by four experienced physicians and two social workers.
% \fixme{avoiding giving advice}

% \subsection{Validation and Usability Test}
% To assess the usability and integration of three AI assistants in PreCare as well as its ability to improve ACP knowledge, readiness and decision-making, we conducted a usability study.

\subsection{Usability study}
To assess the usability and integration of three AI assistants in PreCare, we conducted a usability study.

\subsubsection{Study Design, Procedure, and Participants}
The study was based on recent ACP website evaluation studies~\cite{Roberts2024Apersonalizedandinteractive}. Participants were asked to complete a consent form, and a demographic survey. They then used PreCare independently without any assistance or interference. After using PreCare, participants completed a System Usability Scale (SUS)~\cite{Bangor2009Determiningwhatindividualsusscoresmean, Sauro2011Practical}, followed by a 20-minute semi-structured interview. After completing the session, the participants received nominal compensation for their participation. The entire study lasted approximately 60 to 90 minutes. 
% The System Usability Scale (SUS) was also included in the post-test

\subsubsection{User Data Security}
\label{sec:user_data_security}
Participant privacy was a major consideration in our research. Participants were identified by user IDs rather than their names. One researcher monitored the user interface remotely, with the participant's consent. If participants revealed sensitive data, the researcher would ask them to replace it with unidentifiable information.

Furthermore, data was stored anonymously. A physician reviewed the anonymous transcripts to ensure the LLM provided accurate and unbiased responses. If the LLM gave inaccurate or biased information, the physician would write down the correct information and pass it to the recruitment researcher. The recruitment researcher did not review the data and instead relayed the corrected information to the corresponding participants via their user ID. 

However, the precautionary measures above were never required.

\subsection{Usability study results}
%\paragraph{\textbf{Usage Statistics}}
Participants used PreCare for a duration ranging from 28 to 62 minutes (mean=40.2, SD=10.5). There were 33 FAQs in Section 2, with participants clicking on an average of 15.25 (SD = 13.84) of them. Additionally, 42\% of users (5/12) asked additional questions, which were answered by the AI assistant based on the database curated by physicians.

%\subsubsection{System Usability} \hfill

In terms of usability, participants rated PreCare to have a System Usability Scale (SUS) score of 80.6 (SD = 16.2), which indicated \textit{excellent} user-friendliness~\cite{Bangor2009Determiningwhatindividualsusscoresmean, Sauro2011Practical} and was above the average score of 68.2 for websites.
%\subsubsection{Recommendation for future usage} \hfill
In terms of how strongly participants would recommend PreCare to others, PreCare received an average rating of 9.2 (SD = 0.94) on a 10-point Likert scale.
%(10 being "I strongly agree to recommend this website to others"), 6 participants chose 10, 2 chose 9, and 4 chose 8. 
%\paragraph{\textbf{Subjective Rating}}
% \begin{enumerate}
%     \item \textbf{}
    
%     \item \textbf{}
    
% \end{enumerate}

\subsubsection{Qualitative feedback} \hfill

The participants reported that the website was \pquotes{intuitive}{P5-6, P8}, and noted that the ACP process was \pquotes{smooth and stress-free}{P5, P12}, and that \pquotes{"the language of conversations was quite gentle and neutral, without forcing me to make any specific choice. The process was gradual and using it felt very comfortable."}{P5}
%They also highlighted the seamless integration of the three AI assistants as \pquotes{highly cohesive}{P3, P10}.

Participants were surprised that later parts of the site \pquotes{remembered what I typed earlier}{P1, P3-5, P11}. They reported that the conversation  \pquotes{provided relevant recommendation of options based on my values}{P8}. Additionally, %some participants particularly appreciated real-time Q\&A function. As P7 noted, 
\pquotes{"the website answered my questions anytime, encouraging me to ask more questions."}{P7}

Regarding future usage, P4 mentioned \emph{"I would like to use PreCare regularly, just like taking a shower. PreCare helps cleanse my mental state."} 

For improvement, participants noted that the ACP process was \pquotes{a little too long}{P1, P3, P8} and expressed a desire to \pquotes{skip steps}{P1, P4}, even though our system did provide such skip buttons on each page. When asked why they did not use them, one participant replied, \pquotes{I feel bad not responding when the website has already answered me}{P1}. This suggests the need for the assistants to balance the conversation duration with user motivation to complete the entire process.

% the in-context design \pquotes{felt like it was helping me to take notes.}{P10}

\section{Comparative User Experience study}
To evaluate the user experience of the AI assistants, we conducted a study comparing PreCare \textit{with} vs. \textit{without} AI assistants.

\subsection{Study Design, Procedure, and Participants}
This study used a within-subjects design and was divided into three parts, with each participant experiencing two conditions: PreCare \textit{with} and \textit{without} AI assistants, in counter-balanced ordering. In the first part, participants experienced Section 1 of the website, answering the first three questions for one condition and the last three for the other condition. In the second part, they experienced Section 2, learning about five medical conditions from one condition, and life-sustaining treatment and artificial nutrition from the other. In the third part, they made an Advance Decision within Section 3 twice, first using one condition and then using the other. Section details are listed in Table~\ref{tab:topics_in_precare}. The order of experiencing both PreCare \textit{with} and \textit{without} assistants was consistent for each participant across all three steps, but counter-balanced between participants. After completing each step, participants provided feedback on specific aspects: in Step 1, they compared which condition better supported the exploration of personal values; in Step 2, they evaluated the completeness of knowledge gained; in Step 3, they assessed their confidence in decision-making. After completing all three steps, they were asked to complete a post-study survey to provide feedback on their experiences and conduct a semi-structured interview. The post-study survey focused on overall preference and rating of the conversation experience with AI assistants. The questions were adapted from metrics
proposed by Abd-Alrazaq et al.~\cite{AbdAlrazaq2020Technicalmetricsusedto} for the technical evaluation of healthcare chatbots in carrying out natural conversations~\cite{li2024beyondthewaitingroom}. 
The participants received nominal compensation for their participation. The entire study lasted approximately 90 minutes. Precautionary measures for user data security, as outlined in Section~\ref{sec:user_data_security}, were implemented but never required.

We recruited 12 participants (5 males and 7 females) with ages ranging from 26 to 73 years (mean=51.5, SD=16.8).

\begin{figure*}
    \centering
    \includegraphics[width=\linewidth]{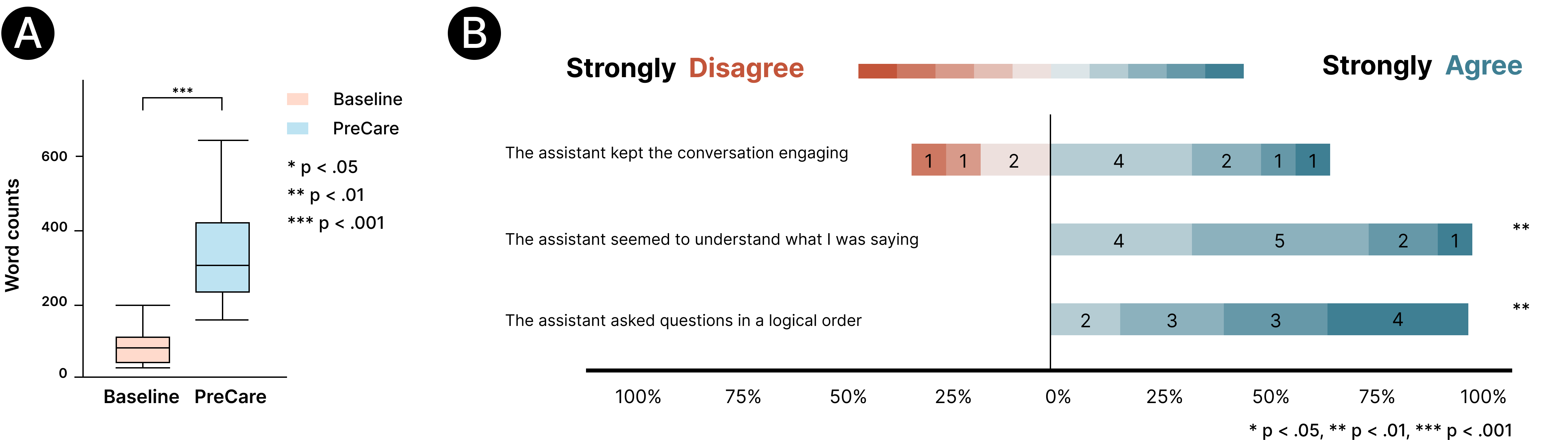}
    \caption{User experience study participants' script analysis and feedback on the AI assistants: (A) the word count in PreCare \textit{with AI assistants} was significantly 
 higher(p<.001) than \textit{without AI assistants}; and (B)Participants’ ratings regarding the perceived quality of the AI assistant according to metrics by Abd-Alrazaq et al.~\cite{AbdAlrazaq2020Technicalmetricsusedto}}
    \label{fig:UX_breadth_chatbot.png_img} 
\end{figure*}
\begin{figure*}
    \centering
\includegraphics[width=\linewidth]{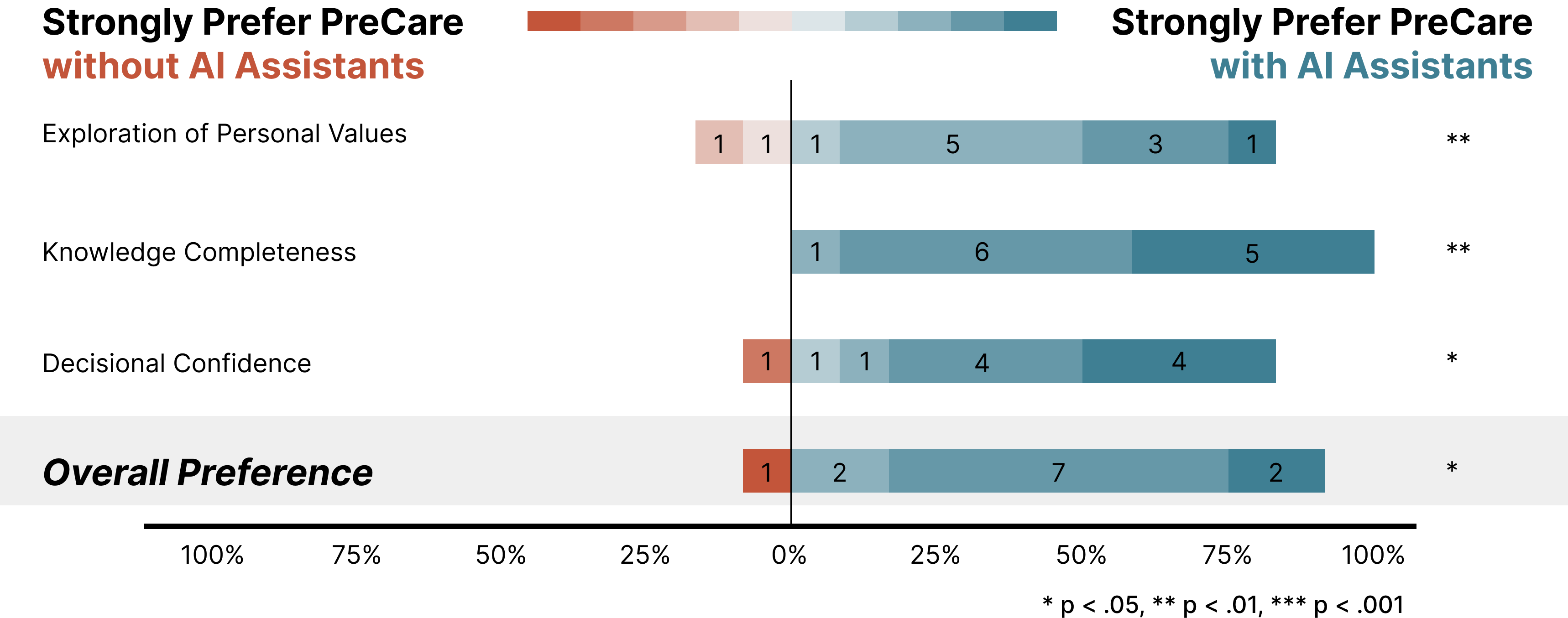}
    \caption{Preference ratings for Advance Care Planning were measured on a 10-point scale, comparing PreCare \textit{with} versus \textit{without AI assistants}. Participants significantly preferred PreCare 
 \emph{with AI assistants} for all aspects, including Exploration of Personal Values(p<.01), Knowledge Completeness(p<.01), Decisional Confidence(p<.05), and Overall(p<.05).}
\label{fig:UX_4metrics_v2.png_img} 
\end{figure*}

\subsection{Conversational Quality of AI assistants}

A two-tailed Wilcoxon signed-rank test was performed for significance analysis.

As shown in Figure~\ref{fig:UX_breadth_chatbot.png_img}(a), the word count of user input in PreCare \textit{with AI assistants} (mean=357.6, SD=179.1) was significantly higher (p<.001) than in PreCare \textit{without AI assistants} design (mean=88.4, SD=52.8). 

Figure~\ref{fig:UX_breadth_chatbot.png_img}(b) shows how participants rated the AI assistants: 67\% of participants felt that the AI assistants were engaging, and all the participants felt that the AI assistants understood what the participants were saying (p<.01), as well as asked questions in a logical order (p<.01).

The results revealed that the AI assistants prompted the users to document more thoughts compared to static questionnaires and could successfully carry out natural conversations.

\subsection{User experience results}

Figure~\ref{fig:UX_4metrics_v2.png_img} shows the 10-point strength-of-preference ratings, using two-tailed Wilcoxon signed-rank tests for significance analysis. Participants significantly preferred PreCare \textit{with AI assistants} for the exploration of their personal values (p<.01), knowledge completeness (p<.01), decisional confidence (p<.05), and 92\% of participants (11/12) preferred PreCare \textit{with AI assistants} for overall experience.

\subsubsection{Exploration of Values.} \hfill

Participants mentioned the AI assistant helped them to \pquotes{consider things they had never thought of before}{P2, P4, P6} through \pquotes{step-by-step guidance}{P1-5, P8} and \pquotes{feedback based on my inputs}{P6-8}. As P4 commented, \emph{"the AI assistant provided feedbacks and directions for further reflection. This often included perspectives I hadn't considered, encouraging deeper thought and helping me better understand my own values."}
Also, the \pquotes{interactive discussions}{P1, P3, P7} helped participants be more willing to document inner thoughts. As P3 mentioned \emph{"I received encouragement and affirmation, which motivated me to continue the conversation and delve deeper into the topic."}

Without AI assistants, participants had to \pquotes{generate all content independently}{P2-3, P11} and was \pquotes{less inclined to write down their thoughts despite engaging in internal reflection}{P7}.

However, 2 participants preferred PreCare \textit{without AI assistants}, as \pquotes{the design without AI assistants allowed me to view all the questions at once, giving me a sense of knowledge for the entire process, which made me feel like I had more opportunity for thorough exploration.}{P2} P10 also mentioned \emph{"the way AI assistants interacted with me makes me want to respond quickly, giving me less time to think before answering. In contrast, the version without AI assistants allows me to have more internal dialogue and reflection."} These findings suggest that while the prompt establishes the discussion goal, users may remain unaware of it due to its implicit nature within the LLM. To address this, we plan to enhance AI assistants by explicitly communicating the discussion goals and endpoints to users. For instance, a progress bar could be integrated into the interface to help inform users of their current progress in the discussion.

\subsubsection{Knowledge Completeness.}\hfill

% The participants all had positive reactions to how the AI assistants presented information on ACP knowledge. 
The AI assistants gave the participants \pquotes{clearer understanding}{P3, P10} of the issues behind treatments, and a better grasp of \pquotes{why making an advance medical decision is important}{P1-2, P5}. The AI assistants encouraged users to engage more actively with the content, as the participants clicked on \pquotes{questions of interest}{P2-5, P11} or those they \pquotes{hadn't considered}{P6-7} of FAQs, deepening their understanding of ACP knowledge. In addition, participants were encouraged to \pquotes{slow down to click the FAQ and use Q\&A function, and reflect on unfamiliar questions and delve deeper into my feelings about medical treatments}{P10}. This interactive approach made the process of acquiring knowledge more engaging and less monotonous. As P7 mentioned, \emph{"when I saw dementia, I felt I understood most of the condition because of my grandma's experience. However, the FAQ highlighted questions such as "Is dementia irreversible?" and "Do dementia patients always need a nasogastric tube?" which helped me better understand the specific scenarios I might need to consider under advance decision."}
For the real-time Q\&A functionality, participants found the AI-generated content appended below the FAQ \pquotes{felt like taking notes}{P3, P11}, which made them \pquotes{wanted to revisit this page to see my notes}{P11}. Additionally, P12 highlighted, \emph{"It was convenient to get accurate answers without searching online."}

Without AI assistants, P6 mentioned, \emph{"it felt like reading a book; it was easy to overlook some details."} Some participants also mentioned that they would \pquotes{assumed I understood knowledge}{P2, p5, P7} after a quick glance and therefore wouldn’t think about them deeply.

\subsubsection{Decisional Confidence.}\hfill

Participants reported that AI assistants \pquotes{quantified the factors}{P6, P8} that need to be considered in the decision-making, which helped them feel more confident in making decisions. The participant mentioned the breadth and depth of considerations before making a decision are more \pquotes{comprehensive}{P1-3, P5-7, P12}. As P6 mentioned, \emph{"having an overview before making the final decision allowed me to review the content discussed earlier, helping me reflect on my values and the knowledge I had learned. This review process made me feel more confident and grounded in my decision-making."} 

Without AI assistants, users had to rely entirely on their own thinking. Forgotten aspects remained overlooked, leading to decisions that might \pquotes{neglect important knowledge or personal values}{P3, P8, P10}. 

However, 1 participant preferred experiences of PreCare \textit{without AI assistants} because \pquotes{AI assistants made me realize that I didn’t fully understand my options, which lowered my confidence and made me feel that I hadn’t thoroughly understood or thought through everything. It gave me a sense of frustration}{P5}. This highlights the need for conversational AI to prioritize empathy (as discussed later) to alleviate users' sense of frustration during the process.

\subsubsection{Overall Preference.}\hfill

Participants reported that compared to PreCare \textit{without AI assistants}, AI assistants in each section \pquotes{guided me to reflect on my values, provided more relevant information, and helped me understand the outcomes behind each option to make a decision}{P2}. As a result, \pquotes{PreCare helped me think more comprehensively and allowed me to approach this serious and important topic with greater detail and care}{P1}. One user who strongly prefer Precare \textit{with AI assistants} even mentioned: \pquotes{It was only after using FAQ and Q\&A function by PreCare (Assistant 2) that I learned the success rate of CPR is lower than its failure rate. If I had only experienced the version without AI assistants design, I wouldn’t have searched for this information and likely would have chosen to accept all treatments. However, with PreCare, I carefully reviewed the FAQs and decided to refuse all treatments}{P7}.

However, one participant (P8) strongly preferred PreCare \textit{without AI assistants}, as \emph{"AI assistants required a lot of thinking. Some follow-up questions left me unsure how to respond, which made me feel stuck. The system without AI assistants did not require as much thinking."}

\section{Discussion, limitations, and future work}

\subsection{Design recommendations}
Drawing on insights from our design process and user studies, we propose design recommendations for AI technologies in related domains.

\subsubsection{Conversational AI with Follow-Up Capabilities for Deeper Reflection} \hfill

Fields requiring user introspection, such as psychological or career counseling, can significantly benefit from conversational AI. Compared to static questionnaires, these AI-driven systems offer dynamic, interactive dialogues that leverage user input to uncover deeper personal values through \emph{feedback} and \emph{follow-up questions}.

Beyond selecting a capable large language model (LLM) with effective prompt design, our experience shows that integrating a grading or evaluation mechanism was essential for follow-up conversation that encourages deeper reflection. This mechanism provides the LLM with a reference for assessing whether a user's response is sufficiently detailed or meets certain criteria before proceeding. If a response falls short, the chatbot can offer tailored guidance, prompting users to elaborate further or meet the necessary conditions for deeper exploration.

% \subsubsection{Conversational AI with follow-up capability for deeper reflection} \hfill

% Fields requiring user reflection, such as psychological and career counseling, can benefit from conversational AI.
% Compared to static questionnaires, these AI assistants offer enhanced conversational capabilities and can utilize user input to explore deeper personal values through \emph{feedback} and \emph{follow-up questions}.

% Beyond selecting a good conversationally capable LLM model with prompt design, we recommend incorporating a grading system into the system for deeper reflection. This system would provide the LLM a reference to assess whether a user’s response is sufficiently detailed or meets certain criteria for proceeding to the next question. If the response falls short, the chatbot can provide tailored guidance to help the user elaborate further or meet the necessary conditions for deeper exploration.

\subsubsection{In-Context Prioritized FAQs \& Real-Time Q\&A for Enhanced Background Knowledge Acquisition} \hfill

For public education systems that convey professional, complex information (e.g., hospital health education, legal advocacy, tax information), we recommend incorporating two key components to accommodate users with varying knowledge levels and support real-time question answering:

\begin{enumerate}
    \item \emph{Prioritized FAQs} allow users to explore topics they find confusing or interesting, providing tailored access to relevant information.
    \item \emph{Real-Time Q\&A} delivers accurate domain-specific answers through Retrieval-Augmented Generation (RAG), enabling immediate responses to user queries.
\end{enumerate}

Additionally, our experience shows that integrating an in-context user interface (UI) design helps reduce visual distractions and simplify content comparison. This approach lowers the barrier to accessing FAQs and posing questions, allowing users to engage seamlessly without disrupting their learning process.

\subsubsection{Personalized Impact Analysis for Decision-Making Systems} \hfill

In decision-making contexts (e.g., real estate purchases, career decisions), we suggest providing personalized impact analysis. Our experience shows that by combining user input with domain knowledge, the system can more effectively help users thoroughly evaluate all aspects of each option, ensuring nothing is overlooked.

We further recommend that the system support both \emph{reviewing} and exploring the \emph{consequences} of different choices based on users' interests or demographics, enabling better comparisons and identification of the option that best fits their needs. 

\subsection{Implications of PreCare for decision-making systems}

Advance care planning (ACP) is a collaborative process where professionals and the general public work together to make decisions that align with both professional considerations and individual preferences. However, effectively conveying professional knowledge to the general public is challenging. To address this, PreCare introduces conversational AI as a vital bridge that not only communicates expert knowledge but also provides real-time Q\&A and performs impact analysis on decisions, enabling choices that balance expertise with personal values.

We believe this workflow has the potential to enhance other medical discussions, such as shared decision-making (SDM). SDM is a process where patients and healthcare providers collaborate to make decisions based on the latest medical evidence, as well as the patients’ preferences and values~\cite{Charles1997Shareddecisionmaking}. For instance, in cancer treatment, physicians may present management options like radiotherapy and chemotherapy, but the final decision rests with the patient~\cite{Bougeard2023Perioperativeshareddecisionmaking}. While current AI interventions in medicine predominantly focus on predicting clinically significant outcomes and making recommendations for physicians~\cite{AbbasgholizadehRahimi2022Applicationofartificialintelligenceinsharedecisionmaking, Hao2024Advancingpatientcenteredshareddecisionmaking}, interactive conversational systems can better engage patients, helping them explore their values, understand trade-offs, and make more informed decisions. Incorporating multidisciplinary knowledge from medical professionals, including physicians, social workers, and nurses, further enriches this process, as demonstrated in our research. However, researchers need to consider some factors. In our study, to ensure Advance Care Planning (ACP) aligns with localized regulations, the system focused on five medical conditions that may impair consciousness and five types of life-sustaining treatment. These topics, being relatively structured, can be organized into a database. Conversely, shared decision-making often requires tailoring to a wide range of diseases, necessitating the integration of specific domain knowledge into large language models (LLMs). When appropriate, demographic data from participants can also serve as input for deep learning models to generate more personalized and effective suggestions.

Beyond medicine, other fields could also adopt AI assistants to help the general population make decisions by integrating expert recommendations with personal preferences. For example, in interior design, designers usually present design options along with their considerations, allowing homeowners to make final decisions; the lawyer presents legal options and potential consequences, while the client decides whether to settle or pursue further appeals. Incorporating conversational AI into such workflows can enhance the process by bridging professional expertise with individual preferences, offering potential benefits across a wide range of domains.
% PreCare在解決的，是專業人員跟一般民眾要合作做出符合專業考量&個人意志的決定。但要將專業知識傳遞給一般民眾很困難，conversational AI就很重要，可以有專業、可以講得很淺白，作為一個橋樑協助。類似領域：醫療shared decision makinig (手術兩種術式，醫生解釋後會給病人自己選擇)。其他可能如室內設計、社工。

\subsection{Implication of PreCare for Thanato-technology}
There has been exciting research in HCI for thanato-technology, which seeks to engage with the experiences of the dying and the bereaved~\cite{Albers2023Deathdyingendoflife} to promote a "good death" for individuals and their loved ones. 
% To date, most thanato-technologies have primarily focused on supporting the bereaved~\cite{Albers2024letstalkaboutdeath}. For example, they aim to reduce grief through reflective experiences of game-based interventions~\cite{LeFevre2024Newunderstandingofloss}. 
To date, most thanato-technologies have primarily focused on emotional support. For example, for the bereaved, study has demonstrated game-based interventions~\cite{LeFevre2024Newunderstandingofloss} reduce grief through reflective experiences; for the dying, achieving emotional clarity by discussing death-related topics with chatbots can help reduce fear and avoidance~\cite{Albers2024letstalkaboutdeath}. 

Our research further emphasizes that making end-of-life care decisions involves more than just emotional clarity. It requires a balanced approach that integrates rational analysis, such as understanding the consequences of medical treatments and the impact on family members, to enable comprehensive discussions. AI assistants could aid users in preparing for their own death by considering expert knowledge, personal values, and the potential impacts of decisions, all tailored to individual needs. 
% Our findings suggest that, in addition to providing emotional support and clarity, it can also integrate professional knowledge to address end-of-life applications effectively.

Thanato-technology encompasses various applications that could benefit from \emph{rational considerations}. For instance, when assisting users in managing digital legacies~\cite{Locasto2011Securityandprivacyconsiderations, Micklitz2013Iherebyleavemyemail, Pfister2017Thiswillcausealotofwork}, the system could not only help think through and communicate intentions and motivations of how they want to preserve their data, and also offer domain-specific knowledge about the implications of their choices, such as cloud storage requires financial costs, while physical storage demands space and maintenance. This could help participants gain a more comprehensive view of the entire issue. 

% Decision-making is an important part of shaping the end of life. Pre iouse work explored a tool that triggers conversations about end-of-life choices and uses value sharing modules to visualize how far positions are apart. Notably, this is a step away from autonomy. 

% There is an unexplored area in confronting people with their mortality.

% What seems to mising is
% Our research 
% 1) reflective technology for the dying and 
% 2) thanatotechnology 

% When it comes to individuals who are approaching death or preparing for end-of-life decisions in advance, most research has concentrated on digital legacies~\cite{Locasto2011Securityandprivacyconsiderations, Micklitz2013Iherebyleavemyemail, Pfister2017Thiswillcausealotofwork} and looking back~\cite{Lee2020Lifereview} to on the life narratives. 

% There is an unexplored area in confronting people with their mortality. Recent studies have shown that achieving emotional clarity by discussing death-related topics with chatbots can help reduce fear and avoidance~\cite{Albers2024letstalkaboutdeath}. 

% Our findings suggest that future thanato-technology could not only provide emotional support and clarity but also incorporate professional knowledge related to end-of-life applications. This insight can be applied not only to decision-making but also to other areas. 

\subsection{ACP technology tailored for different age groups}
Current ACP approaches primarily emphasize medical care decisions, often neglecting the underlying feelings, thoughts, and values that are equally important for a holistic discussion~\cite{Groebe2019Howtotalkaboutattitudes}. PreCare emphasizes the equal importance of decision-making and the process leading to it, highlighting the advantages of AI assistants in exploring personal values, acquiring medical knowledge, and performing impact analyses. Thus, we further found that users from different age groups prioritize distinct aspects of PreCare due to their unique life experiences:

% \fixme{need quotes}
\begin{enumerate}
    \item \textbf{Younger users: value the consultation process.} Participants noted two key benefits: 1) Reassessing their values at different stages of life through PreCare. Participants mentioned that using PreCare over time helped them better understand their \pquotes{core values}{Usability: P4, P7; Comparative: P7-9, P12}, including which experiences shaped their values \pquotes{at different stages of life}{Usability: P4} and which values \pquotes{remained constant overtime}{Comparative: P7, P12}; 2) Acquiring medical knowledge, which helps to \pquotes{enhance basic understanding of medical scenarios and consequences}{Comparative: P1, P3-4, P7}, and educates them \pquotes{stay healthy}{Comparative: P1} and holds the potential to reduce unnecessary medical interventions and encourages more informed decisions \pquotes{when family members are ill}{Usability: P6-8}  (e.g., avoiding insistence on futile treatments). As P1 mentioned in the comparative study, \emph{knowing that people in a vegetative state may stay in bed for a long time, it reminds me to drive more cautiously to avoid car accidents.}
    \item \textbf{Older users: focus on ensuring that advance decisions align precisely with their values.} 
    Participants mentioned that PreCare not only generates an advance decision but also provides \pquotes{a conversation summary}{Usability: P1-2, P11}. This summary has the potential to help physicians and family members better understand \pquotes{why I made such decisions}{Usability: P11}. Currently, if a patient loses consciousness, the only reference is their advance decision. However, when family members disagree with the patient's choice, doctors must rely solely on the advance decision to persuade them. A conversation summary could bridge this gap by offering deeper insights into the patient’s values and reasoning, helping both doctors and family members trust that the decision aligns with the patient’s true needs.
\end{enumerate}

% In addition, several ACP professionals expected PreCare to provide a summary of users' conversations. Such a summary would be particularly valuable for palliative doctors responsible for aligning end-of-life treatment with patients' wishes. 

These differences highlight the future focus to design tailored systems for different age groups. For younger users, the system could take an educational approach, emphasizing value exploration and medical knowledge acquisition. To enhance engagement for the young, more interactive elements could be provided, such as gamification experiences~\cite{Steenstra2024Engagingandentertaining}. In addition, content should be tailored for young users, especially those without family illness experience, by providing instructional videos on treatment options. The system could initially guide users to consider more common scenarios (e.g. vegetative state) for the young, before expanding to other contexts (e.g. dementia). 

For older users, the system could prioritize impact analysis to ensure the selected options truly reflect their preferences, while also documenting the consultation content. To improve accessibility, it could focus on natural voice interactions~\cite{Yan2024Understandingolderpeoplesvoiceinteractions} for ease of use.

% \fixme{natural language and face to face }

% 因為死亡相對比較遠，如果不是實驗，對於AD比較沒有動機。

% In addition, participants reponseded 探索跟醫學

% 重視的是探索跟醫學知識，比較偏教育的感覺。

% Everyone should consider advance care planning, regardless of age or health status~\cite{AdvancecareplanningAustralia2024}. Younger individuals tend to have a higher acceptance of technology, but may feel that end-of-life planning is not immediately relevant to them. In contrast, older adults, often with chronic conditions or experiences of losing loved ones, tend to show more interest in ACP, but may be less familiar with technology. Therefore, it is important to design for different age groups, taking into account their varying needs and levels of familiarity with technology.

% For younger audiences, a more interactive or gamified design could better capture their interest and engagement~\cite{Steenstra2024Engagingandentertaining}. %VR game or Avatar work
% For older adults, face-to-face interactions are often preferred over conversations with a chatbot, and they may be less familiar with text-based interactions~\cite{Goot2020Exploringagedifferencesinmotivations}. In our Usability Study, we also found a moderate negative correlation between patient age and system usability ratings, though it was not statistically significant (r=-0.52; P=.083). While all participants were able to use the system independently, to make it more accessible for older users, we are currently designing natural voice interactions that eliminate the need for typing.
\subsection{A more personalized AI assistant}
% \fixme{need quotes}
% \fixme{more empathy for impact analysis with pquotes}
While this work demonstrated an approach that utilizes user input to facilitate informed decision-making compared to the static questionnaires on current websites, there is potential to leverage even more user input to enhance engagement.

For example, many participants emphasized that AI assistants should focus on empathy. Sometimes feedbacks are \pquotes{too formal}{Usability: P4-5, P10, P12; Comparative: P2, P5} and less human-like. As P12 from usability study stated \emph{"AI assistants sometimes said, "Thank you for your response." While polite, this phrase is uncommon in natural human conversations."} In addition, we found some participants desired \pquotes{more delay time}{Usability: P7; Comparative study: P5, P9} in AI responses. As P5 explained, \emph{"with more delay time, it feels like the system is taking more time to think about my input, which makes me feel my thoughts are being taken seriously."} 

Surprisingly, while we assumed that discussions about one’s mortality might provoke fear and avoidance~\cite{Albers2024letstalkaboutdeath}, some users instead prioritized efficiency. These participants felt the chatbot should \pquotes{avoid excessive expressions of care}{Usability: P3; Comparative: P8}, preferring content that is \pquotes{precise}{Comparative: P8} and to the point. For these participants, they wanted \pquotes{shorter response times}{P3}, enabling them to iterate their thoughts more frequently.

These findings suggest that, beyond considering age differences, systems should allow users to customize parameters such as tone, empathy level of contents, and response delay. Personalization options can ensure that the system aligns with diverse user needs, balancing empathy and efficiency to provide an optimal experience for decision-making.

To make content more empathetic, we are currently using audio transcripts from an experienced ACP physician to train our model, aiming to integrate the physician's methods of expressing empathy into our system.

\subsection{Limitations and Future Work}
 % agent有用。但對於實際臨床效果要用randomized control 效果來確認

First, our focus is on designing AI agents to facilitate the process of a self-guided ACP website. This study provides the first empirical evidence of AI's role in advance care planning, demonstrating its ability to support participants in making more informed decisions. In the future, this approach can be expanded to clinical settings, broaden its user coverage, and be validated through randomized controlled trials.
For example, the user pool can expand to individuals with a clear prognosis or circumstances indicating imminent death. They may face unique challenges, such as limited ability to interact with others or technology~\cite{Massimi2011Mattersoflifeanddeath}. Advance care planning for these patients prioritizes both empathetic and efficient decision-making processes, supported by a user-friendly interface. In addition, the decisions or experiences encompass could not only medical treatments but also considerations regarding legacies and memorials~\cite{Albers2023Deathdyingendoflife}. 
% the study intervention requires further validation through a randomized controlled trial and more participants to achieve a higher level of evidence. In addition, we plan to follow up on the status of the patients. Potential outcomes include behavior change and ACP/AD actions from previous ACP study~\cite{Sudore2014Anovelwebsite}. 

Second, our current version supports text input through typing or speech-to-text functionality. We plan to expand input options to accommodate diverse user preferences. Currently, we are exploring the use of natural language interfaces to enhance user experience.
% Second, our current version support type or use speech-to-text for typing texts. We would expand input來滿足不同user preference。we are currently explore natural language interface..

% However, for individuals unfamiliar with text input or those whose medical condition leaves them conscious but unable to care for themselves, our system would be unusable. 

% Additionally, participants in both the Usability Study and the Comparative User Experience Study had higher education levels, as shown in Table\ref{tab:demographic}, highlighting the need for further research on interactive design for users with diverse educational backgrounds. Therefore, further user testing with more diverse populations, gathering qualitative feedback, or administering additional surveys for deeper insights is necessary. 

% Third, our participants represent a healthy general population across various age groups. However, 

% In addition, the systems may involve incorporating three main themes identified in previous research~\cite{Albers2023Deathdyingendoflife}—"preparation," "retrospection," and "decision-making"—simultaneously.
%It will be important to analyze usability metrics in more detail to identify specific areas for improvement. 

Third, our current version focuses primarily on one user, the decision maker, but also provides information and tips to encourage discussions with family members, caregivers, or surrogate decision-makers. However, different roles may require tailored information or interaction. As Foong et al.~\cite{Foong2024Designingforcaregiver} highlighted, caregivers often need more support in understanding their own values. 
Future work involves developing decision-making systems for group consultations and support systems for surrogate decision-makers.

Finally, large language models (LLMs) offer strong conversational capabilities and can provide guidance tailored to specific designs. As AI technology continues to improve, it is crucial to ensure that these systems do not produce outputs that are inaccurate, incomplete, incorrect, or offensive when introducing new LLMs into systems~\cite{OpenAI2023Termsofuse}.

% we utilized the OpenAI API. Due to the probabilistic nature of LLMs, they may provide inaccurate, incomplete, incorrect, or offensive Outputs~\cite{OpenAI2023Termsofuse}. Therefore, in our research, we implemented a human-in-the-loop process to ensure user data privacy and anonymity. In the future, we plan to train our own model for the API and integrate an additional model, acting as the human role in the research, to enhance security and content verification. Many AI research efforts focus on safety and alignment\cite{Ji2023AIAlignment}. we will also incorporate these relevant technologies to ensure safety before deploying the system to a larger audience.

% 很多AI research在safete, alignment。我們如果deploy給更多人用會將技術放進來

%大眾的想法 跟我發現的有什麼不一樣
%point out負面的地方
% AI assistant我從中學到什麼、AI用在這個上面
% AI technology 有什麼可以改進for future ACP
% 現在第一代可以有introspection, 互動、consistent quality。但還缺乏什麼。
% address ethical issues。怎麼確保。因為我們是introspection。但如果有guiding，要小心不要Introduce bias。但其實這個在與真人互動也會有這樣的問題，真人也是有bias但努力盡量不要有bias。我們AI assistant也是努力盡量保持客觀。
% \subsection{More functionalities to facilitate ACP}
% We are exploring several functionalities to enhance user experience and support decision-making, and we are currently testing their effectiveness.

%更精準的分類
%

% \subsection{For elders}
% \subsubsection{Voice Output} 
% design for aging need improve
% voice conversation
% 影片取代文字...

%有更多voice input & output功能。比text-based還好。更natural voice interaction 

% \subsection{Future work}

% \subsection{Limitations}

\section{Conclusion}
In this paper, we presented the first exploration of designing AI assistants for Advance Care Planning (ACP), in collaboration with ACP professionals to combine the key benefits of clinical consultation with the accessibility of online ACP platforms.
Through an iterative design process and a series of formative and summative studies involving medical professionals and 38 end-users, our results demonstrated the feasibility of LLM-based AI assistants in enhancing introspection on personal values related to end-of-life issues, increasing knowledge acquisition, and encouraging thoughtful consideration of the consequences of decisions.
Based on the study, we provided a set of design recommendations that are likely applicable to future work and AI-enabled decision-making systems.
% We conducted a usability study that confirmed the system is easy to use and demonstrated significant improvements in participants' knowledge of ACP, readiness for decision-making, and a significant reduction in the decisional conflict scale. All participants would recommend PreCare to the target group.

% \input{relatedwork}
% \input{method}
% \input{experiment}
% \input{limitation}
% \input{conclusion}
% \begin{acks}
% We thank all participants for their valuable feedback, and Cindy Lin, Mireille Tan, Yu Chen, and Pin-Chun Lu for their kind suggestions. We also gratefully acknowledge the Hospice Foundation of Taiwan and the Academy of Humanities and Innovation at Taipei City Hospital for providing ACP-related resources.
% \end{acks}
{\small
\bibliographystyle{ieee_fullname}
\bibliography{paper}
}    
\appendix
\label{sec:appendix}
\onecolumn

\section{Demographics}

\begin{table}[ht]
\caption{The demographics of our participants (N=38) in three studies.}
\label{tab:demographic}
\resizebox{\textwidth}{!}{
\begin{tabular}{llrrr}
\hline
                &                                 & \begin{tabular}[c]{@{}r@{}}User Experience Evaluations \\ with the ACP Website\end{tabular} & Usability study & \begin{tabular}[c]{@{}r@{}}Comparative\\ User experience study\end{tabular} \\ \hline
Gender          & Male                            & 7                                                                                           & 5               & 5                                                                           \\
                & Female                          & 7                                                                                           & 7               & 7                                                                           \\ \hline
Age             & 18-40                           & 9                                                                                           & 4               & 3                                                                           \\
                & 40-60                           & 3                                                                                           & 4               & 6                                                                           \\
                & \textgreater{}60                & 2                                                                                           & 4               & 3                                                                           \\ \hline
Education Level & High school                     & 5                                                                                           & 1               & 1                                                                           \\
                & University Bachelor’s degree    & 5                                                                                           & 6               & 8                                                                           \\
                & Graduate or professional degree & 4                                                                                           & 5               & 3                                                                           \\ \hline
\end{tabular}
}
\end{table}

\section{Prompt design for AI assistant}

% \usepackage{multirow}
\begin{table}[ht]

\caption{The script provided to AI assistant 1 in PreCare. The script was iteratively designed with social workers and end users to avoid the introduction of biased feedback or questions.}
\label{tab:promptdesign1}
\resizebox{\textwidth}{!}{
\begin{tabular}{|cl}
\hline
\multicolumn{2}{|c|}{Prompt Design}                                                                                                                                                                                                                                                                                                                                                                                                                                                                                                                                                                                                                                                            \\ \hline
\multicolumn{1}{|c|}{\multirow{6}{*}{\begin{tabular}[c]{@{}c@{}}Behavior \\ Instructions\end{tabular}}} & \multicolumn{1}{l|}{You are an assistant helping the user think about questions to explore personal values toward end-of-life issues}                                                                                                                                                                                                                                                                                                                                                                                                                                                \\ \cline{2-2} 
\multicolumn{1}{|c|}{}                                                                                  & \multicolumn{1}{l|}{\begin{tabular}[c]{@{}l@{}}If the topic goal is related to end-of-life values and decisions, you should wait until after at least 2 rounds of discussion\\ before asking about specific end-of-life scenarios\end{tabular}}                                                                                                                                                                                                                                                                                                                                      \\ \cline{2-2} 
\multicolumn{1}{|c|}{}                                                                                  & \multicolumn{1}{l|}{The current topic goal: \{goals{[}page\_number{]}\}.}                                                                                                                                                                                                                                                                                                                                                                                                                                                                                                            \\ \cline{2-2} 
\multicolumn{1}{|c|}{}                                                                                  & \multicolumn{1}{l|}{You can ask questions chosen from but not limited to the following examples: \{questions{[}page\_number{]}\}}                                                                                                                                                                                                                                                                                                                                                                                                                                                    \\ \cline{2-2} 
\multicolumn{1}{|c|}{}                                                                                  & \multicolumn{1}{l|}{You should also contain feedback to user's input in your response.}                                                                                                                                                                                                                                                                                                                                                                                                                                                                                              \\ \cline{2-2} 
\multicolumn{1}{|c|}{}                                                                                  & \multicolumn{1}{l|}{Your feedback should incorporate the topic goal and the user’s content to provide empathy.}                                                                                                                                                                                                                                                                                                                                                                                                                                                                      \\ \hline
\multicolumn{1}{|c|}{\multirow{7}{*}{\begin{tabular}[c]{@{}c@{}}Basic \\ Instructions\end{tabular}}}    & \multicolumn{1}{l|}{Do not make medical recommendations to the user.}                                                                                                                                                                                                                                                                                                                                                                                                                                                                                                                \\ \cline{2-2} 
\multicolumn{1}{|c|}{}                                                                                  & \multicolumn{1}{l|}{You must keep your response within 5 sentences and at most 200 words.}                                                                                                                                                                                                                                                                                                                                                                                                                                                                                           \\ \cline{2-2} 
\multicolumn{1}{|c|}{}                                                                                  & \multicolumn{1}{l|}{\begin{tabular}[c]{@{}l@{}}The content of the question you ask should not overlap with content \\ that the user has already input in past conversation memory.\end{tabular}}                                                                                                                                                                                                                                                                                                                                                                                     \\ \cline{2-2} 
\multicolumn{1}{|c|}{}                                                                                  & \multicolumn{1}{l|}{You must not ask questions that have already been asked.}                                                                                                                                                                                                                                                                                                                                                                                                                                                                                                        \\ \cline{2-2} 
\multicolumn{1}{|c|}{}                                                                                  & Ask one question at a time                                                                                                                                                                                                                                                                                                                                                                                                                                                                                                                                                           \\ \cline{2-2} 
\multicolumn{1}{|c|}{}                                                                                  & \multicolumn{1}{l|}{\begin{tabular}[c]{@{}l@{}}The user's input is evaluated with following criteria: \\ 1. Relevance: How well does the user's answer address the topic or question at hand?\\  Is it related to the context of the dialogue? \\ 2. Appropriateness: How appropriate is the user's answer is based on the preceding turn? \\  Are ideas expressed concisely and without ambiguity? \\ 3. Content: Does the user show deep consideration of the subject? \\  Is the answer informative, containing various entities or descriptive or emotional words?\end{tabular}} \\ \cline{2-2} 
\multicolumn{1}{|c|}{}                                                                                  & \multicolumn{1}{l|}{You can reference to the evaluation scores and ask follow-up questions if the user's response is not satisfactory.}                                                                                                                                                                                                                                                                                                                                                                                                                                              \\ \hline
\multicolumn{2}{|c|}{Goals for each topic}                                                                                                                                                                                                                                                                                                                                                                                                                                                                                                                                                                                                                                                     \\ \hline
\multicolumn{1}{|c|}{Topic 1}                                                                           & \multicolumn{1}{l|}{Discuss things that make you happy. Reflect on the goals and wishes that matter most when facing the end of life.}                                                                                                                                                                                                                                                                                                                                                                                                                                               \\ \hline
\multicolumn{1}{|c|}{Topic 2}                                                                           & \multicolumn{1}{l|}{Discuss your attitude towards illness and reflect on your values regarding quality and the meaning of life.}                                                                                                                                                                                                                                                                                                                                                                                                                                                     \\ \hline
\multicolumn{1}{|c|}{Topic 3}                                                                           & \multicolumn{1}{l|}{\begin{tabular}[c]{@{}l@{}}Connect with your own or a loved one's experience of illness, and reflect \\ on your values and attitude towards the end of life.\end{tabular}}                                                                                                                                                                                                                                                                                                                                                                                       \\ \hline
\multicolumn{1}{|c|}{Topic 4}                                                                           & \multicolumn{1}{l|}{\begin{tabular}[c]{@{}l@{}}Imagine the potential scenarios where you might be unable to care for yourself, \\ and discuss who or what could help you in such situations.\end{tabular}}                                                                                                                                                                                                                                                                                                                                                                           \\ \hline
\multicolumn{1}{|c|}{Topic 5}                                                                           & \multicolumn{1}{l|}{\begin{tabular}[c]{@{}l@{}}Discuss the choice of a location for the end of life, and reflect on the people \\ and things you want around you when that time comes.\end{tabular}}                                                                                                                                                                                                                                                                                                                                                                                 \\ \hline
\multicolumn{1}{|c|}{Topic 6}                                                                           & \multicolumn{1}{l|}{Discuss burial preferences, ensuring that after your death, others understand your wishes and preferences.}                                                                                                                                                                                                                                                                                                                                                                                                                                                      \\ \hline
\end{tabular}
}
\end{table}

\newpage

\begin{table}[ht]

\caption{The script provided to AI assistant 3 in PreCare. The script was iteratively designed with social workers, physicians and end users to avoid the introduction of biased feedback or questions.}
\label{tab:promptdesign3}
\resizebox{\textwidth}{!}{
\begin{tabular}{|cl|}
\hline
\multicolumn{2}{|c|}{Prompt Design}                                                                                                                                                                                                                                                                                                                             \\ \hline
\multicolumn{1}{|c|}{\multirow{7}{*}{\begin{tabular}[c]{@{}c@{}}Behavior \\ Instructions\end{tabular}}} & You are an assistant helping the user think about the important aspects before making an Advance Decision.                                                                                                                                                    \\ \cline{2-2} 
\multicolumn{1}{|c|}{}                                                                                  & Discuss preferences in the event of losing consciousness due to a serious illness.                                                                                                                                                                    \\ \cline{2-2} 
\multicolumn{1}{|c|}{}                                                                                  & First, ask the user if, at the end of life, they would prefer to live as long as possible or focus on comfort rather than extending life.                                                                                                             \\ \cline{2-2} 
\multicolumn{1}{|c|}{}                                                                                  & Second, help the user in reviewing the circumstances under which an Advance Decision will be applied.                                                                                                                                                 \\ \cline{2-2} 
\multicolumn{1}{|c|}{}                                                                                  & \begin{tabular}[c]{@{}l@{}}Third, clarify the reasons behind the 4 choices, including \\ "benefits and side effects", "quality of life", "medical expenses", "real-life stories", "caregivers' responsibilities", and "long-term impact".\end{tabular} \\ \cline{2-2} 
\multicolumn{1}{|c|}{}                                                                                  & The 4 choices include: "try all treatments," "do a trial," "reject all treatments," or "delegate decisions to my proxy".                                                                                                                                   \\ \cline{2-2} 
\multicolumn{1}{|c|}{}                                                                                  & You can use materials from \{knowledge\} to help the user think through the aspects of these choices.                                                                                                                                                            \\ \hline
\multicolumn{1}{|c|}{\begin{tabular}[c]{@{}c@{}}Basic \\ Instructions\end{tabular}}                     & The same as those for Assistant 1.                                                                                                                                                                                                                     \\ \hline
\end{tabular}
}
\end{table}

\end{document}